\pdfoutput=1
\documentclass[11pt,twoside,a4paper,cmspaper,final,collab]{cms-tdr}
\def\svnVersion{4b19e58-D}\def\svnDate{2021/08/09}\def\cmsCernNoTag{CERN-EP-2021-045}\def\cmsCernDate{\today}\def\cmsMessage{"Published in the European Physical Journal C as \href{http://dx.doi.org/10.1140/epjc/s10052-021-09472-3}{\doi{10.1140/epjc/s10052-021-09472-3}.}"}
\begin{document}\cmsNoteHeader{HIG-20-017}

\newcommand{\mll}{\ensuremath{\mathrm{m}_{\ell\ell}}\xspace}
\newcommand{\mlll}{\ensuremath{m_{\ell\ell\ell}}\xspace}
\newcommand{\jet}{\ensuremath{\mathrm{j}}}
\newcommand{\mjj}{\ensuremath{m_{\jet\jet}}\xspace}
\newcommand{\detajj}{\ensuremath{\abs{\Delta\eta_{\jet\jet}}}\xspace}
\newcommand{\WW}{\ensuremath{\PW^\pm\PW^\pm}\xspace}
\newcommand{\WZ}{\ensuremath{\PW\PZ}\xspace}
\newcommand{\tZq}{\ensuremath{\PQt\PZ\Pq}\xspace}
\newcommand{\tVx}{\ensuremath{\PQt\PV\mathrm{x}}\xspace}
\newcommand{\zepmax}{\ensuremath{\max(z_{\ell}^{*})}\xspace}
\newlength\cmsTabSkip\setlength{\cmsTabSkip}{1ex}
\newcommand{\PHpmpm}{\ensuremath{\PH^{\pm\pm}}}
\ifthenelse{\boolean{cms@external}}{\providecommand{\cmsLeft}{upper\xspace}}{\providecommand{\cmsLeft}{left\xspace}}
\ifthenelse{\boolean{cms@external}}{\providecommand{\cmsRight}{lower\xspace}}{\providecommand{\cmsRight}{right\xspace}}

\cmsNoteHeader{HIG-20-017}

\title{Search for charged Higgs bosons produced in vector boson fusion processes and decaying into vector boson pairs in proton-proton collisions at  \texorpdfstring{$\sqrt{s} = 13\TeV$}{sqrt(s) = 13 TeV}}
\titlerunning{Search for charged Higgs bosons produced in vector boson fusion processes and decaying into vector boson pairs}

\author*[inst1]{CMS experiment}

\date{\today}

\abstract{
A search for charged Higgs bosons produced in vector boson fusion processes and decaying into vector bosons, using proton-proton collisions at $\sqrt{s}=13\TeV$ at the LHC, is reported. The data sample corresponds to an integrated luminosity of 137\fbinv collected with the CMS detector.  Events are selected by requiring two or three electrons or muons,  moderate missing transverse momentum, and two jets with a large rapidity separation  and a large dijet mass. No excess of events with respect to the standard model  background predictions is observed. Model independent upper limits at 95\%  confidence level are reported on the product of the cross section and branching fraction for vector boson fusion production  of charged Higgs bosons as a function of mass, from 200 to 3000\GeV.  The results are interpreted in the context of the Georgi--Machacek model.}

\hypersetup{%
pdfauthor={CMS Collaboration},%
pdftitle={Search for charged Higgs bosons produced in vector boson fusion processes and decaying into vector boson pairs in proton-proton collisions at sqrts = 13 TeV},%
pdfsubject={CMS},%
pdfkeywords={CMS, diboson, electroweak, charged higgs}}

\maketitle

\section{Introduction}
\label{sec:introduction}

The discovery~\cite{AtlasPaperCombination,CMSPaperCombination, CMSPaperCombination2} of a Higgs 
boson~\cite{PhysRevLett.13.321,Higgs:1964ia,PhysRevLett.13.508,PhysRevLett.13.585,PhysRev.145.1156,PhysRev.155.1554} 
at the CERN LHC marks an important milestone in the exploration of the electroweak (EW) sector 
of the standard model (SM) of particle physics. Measurements of vector boson scattering (VBS) 
processes at the LHC may reveal hints for extensions of the SM. In particular, extended Higgs 
sectors with additional SU(2) doublets~\cite{PhysRevD.8.1226,FAYET197414, Craig2012, Branco20121} 
or triplets~\cite{KONETSCHNY1977433, MAGG198061, PhysRevD.22.2860,PhysRevD.22.2227,CE1,CE2} 
introduce couplings of gauge bosons to heavy neutral or charged Higgs bosons with specific signatures like singly or doubly charged Higgs boson decays to $\WZ$ boson pairs or same-sign $\WW$ boson pairs, respectively. 

At the LHC, interactions from VBS are characterized by the presence of 
two gauge bosons in association with two forward jets with a large pseudorapidity 
separation ($\detajj$) and a large dijet invariant mass ($\mjj$). An excess of events with respect to the SM predictions could indicate the presence of new resonances, such as singly or doubly charged Higgs bosons. 
Extended Higgs sectors with additional SU(2) isotriplet scalars give rise to charged Higgs bosons with couplings to $\PW$ and $\PZ$ bosons 
at the tree-level~\cite{CE2}. Specifically, the Georgi--Machacek (GM) 
model~\cite{GEORGI1985463, CE1}, with both real and complex triplets, preserves a 
global symmetry SU(2)$_\mathrm{L}\times$SU(2)$_\mathrm{R}$, which is 
broken by the Higgs vacuum expectation value to the diagonal subgroup 
SU(2)$_{\mathrm{L}+\mathrm{R}}$. Thus, the tree-level ratio of the 
$\PW$ and $\PZ$ boson masses is protected against large radiative corrections. 
In this model, singly (doubly) charged Higgs bosons that decay to $\PW$ and $\PZ$ bosons (same-sign $\PW$ boson pairs) are produced via vector boson fusion (VBF).

\begin{figure}[htbp]
\centering
\includegraphics[width=0.49\textwidth]{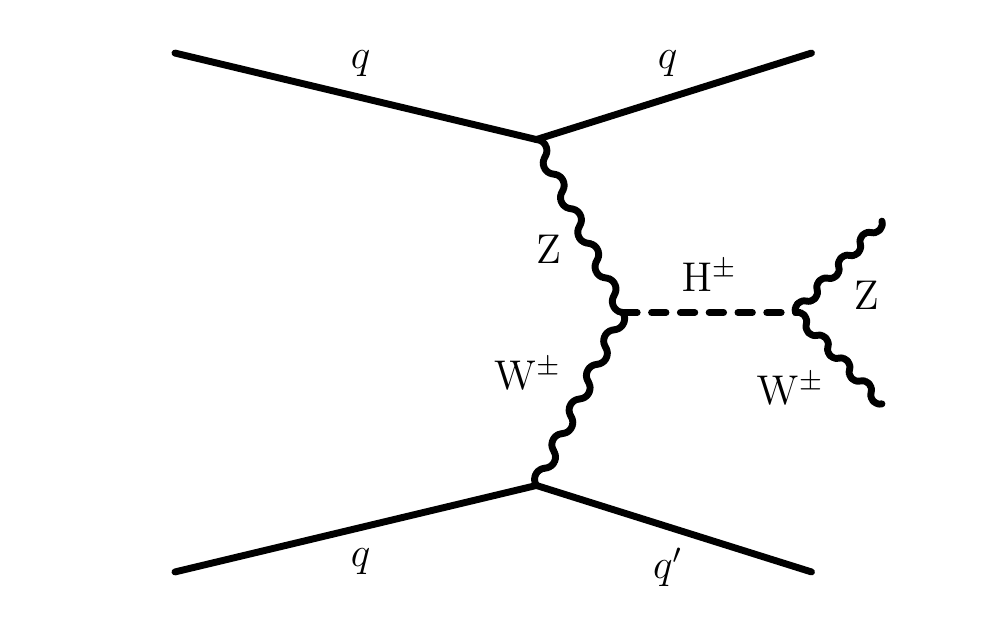}
\includegraphics[width=0.49\textwidth]{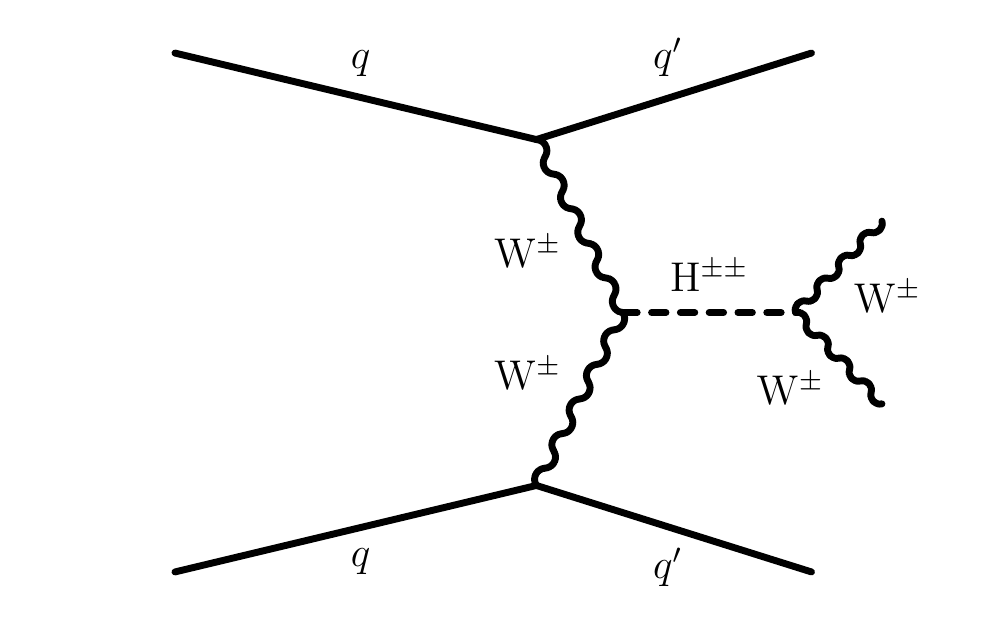}
\caption{Examples of Feynman diagrams showing the production of singly (\cmsLeft) and doubly (\cmsRight) charged Higgs bosons via VBF.}
\label{fig:feynman2}
\end{figure}

The charged Higgs bosons $\PHpm$ and $\PHpmpm$ in the GM model are degenerate 
in mass (denoted as $m_{\PH_{5}}$) at tree level and transform as a quintuplet 
under the SU(2)$_{\mathrm{L}+\mathrm{R}}$ symmetry. The $\PHpm$ and $\PHpmpm$ bosons 
are also collectively referred to as $\PH_5$ in the context of the GM model. Production and decays of the $\PH_5$ states depend on the two parameters 
$m_{\PH_{5}}$ and $s_{\PH}$, where $s_{\PH}^2$ characterizes 
the fraction of the $\PW$ boson mass squared generated by the vacuum expectation 
value of the triplet fields. The $\PH_5$ states are fermiophobic and are assumed to decay to vector boson pairs with branching fraction of 100\%~\cite{Zaro:2002500}. Figure~\ref{fig:feynman2} shows representative Feynman 
diagrams for the production and decay of the charged Higgs bosons. 
There are additional charged Higgs bosons $\PHpm$ predicted in the GM model that 
transform as a triplet under the SU(2)$_{\mathrm{L}+\mathrm{R}}$ symmetry. 
These $\PHpm$ bosons have only fermionic couplings and are not considered here.  

This paper presents a search for $\PHpm$ and $\PHpmpm$ that are produced via 
VBF and decay to $\WZ$ and $\WW$ boson pairs, respectively, using proton-proton ($\Pp\Pp$) collisions 
at $\sqrt{s}=13\TeV$. The data sample corresponds to an integrated luminosity of 
$137 \pm 2\fbinv$~\cite{CMS-PAS-LUM-17-001,CMS-PAS-LUM-17-004,CMS-PAS-LUM-18-002}, 
collected with the CMS detector~\cite{Chatrchyan:2008aa} in three separate 
LHC operating periods during 2016, 2017, and 2018. The three data sets are 
analyzed independently, with appropriate calibrations and corrections, 
to account for the various LHC running conditions and the performance of the CMS detector. 

The $\WW$ and $\WZ$ channels are simultaneously studied by performing a binned 
maximum-likelihood fit of distributions sensitive to these processes, 
following the methods described in Ref.~\cite{Sirunyan:2020gyx}. The searches for 
$\PHpm$ and $\PHpmpm$ are performed in the leptonic decay modes 
$\PW^\pm\PZ\to\ell^\pm\PGn\ell'^\pm\ell'^\mp$ and $\WW \to \ell^\pm\PGn\ell'^\pm\PGn$, where $\ell, \ell' = \Pe$, 
$\PGm$. Candidate events contain either two identified leptons of the same charge or three identified charged leptons with the total charge of $\pm$1, moderate missing transverse momentum ($\ptmiss$), and two jets with large values of $\detajj$ and $\mjj$. 

Model independent upper limits at 95\% confidence level ($\CL$) are reported on the product of the cross section and branching fraction for vector boson fusion production  of the $\PHpm$ and $\PHpmpm$ bosons individually. The results are also interpreted in the context of the GM model including the simultaneous contributions of the $\PHpm$ and $\PHpmpm$ bosons. Searches for charged Higgs bosons in these topologies have been performed by the 
CMS Collaboration at $13\TeV$ using the data sample collected during 2016~\cite{Sirunyan:2017sbn,Sirunyan:2017ret,Sirunyan:2019ksz}. 
The ATLAS and CMS Collaborations have also set constraints on the GM model by performing 
searches for charged Higgs bosons in semileptonic final states at $8\TeV$~\cite{PhysRevLett.114.231801} 
and $13\TeV$~\cite{Sirunyan:2019der}, respectively.

\section{The CMS detector}
\label{sec:cms}

The central feature of the CMS apparatus is a superconducting solenoid of 
6\unit{m} internal diameter, providing a magnetic field of 3.8\unit{T}. Within 
the solenoid volume are a silicon pixel and strip tracker, a lead-tungstate 
crystal electromagnetic calorimeter (ECAL), and a brass and scintillator hadron 
calorimeter, each composed of a barrel and two endcap sections. Forward 
calorimeters extend the $\eta$ coverage provided by the barrel 
and endcap detectors up to $\abs{\eta}<5$. Muons are detected in gas-ionization chambers embedded in 
the steel magnetic flux-return yoke outside the solenoid. A more detailed description 
of the CMS detector, together with a definition of the coordinate system and 
the relevant kinematic variables, is reported in Ref.~\cite{Chatrchyan:2008aa}. 
Events of interest are selected using a two-tiered trigger 
system~\cite{Khachatryan:2016bia}. The first level, composed of custom hardware 
processors, uses information from the calorimeters and muon detectors to select events 
at a rate of around 100\unit{kHz} within a fixed latency of 4\mus. 
The second level, known as the high-level trigger, consists of a farm of 
processors running a version of the full event reconstruction software optimized 
for fast processing, and reduces the event rate to around 1\unit{kHz} before data storage.

\section{Signal and background simulation}
\label{sec:samples}

\begin{figure*}[htb]
\centering
\includegraphics[width=0.49\textwidth]{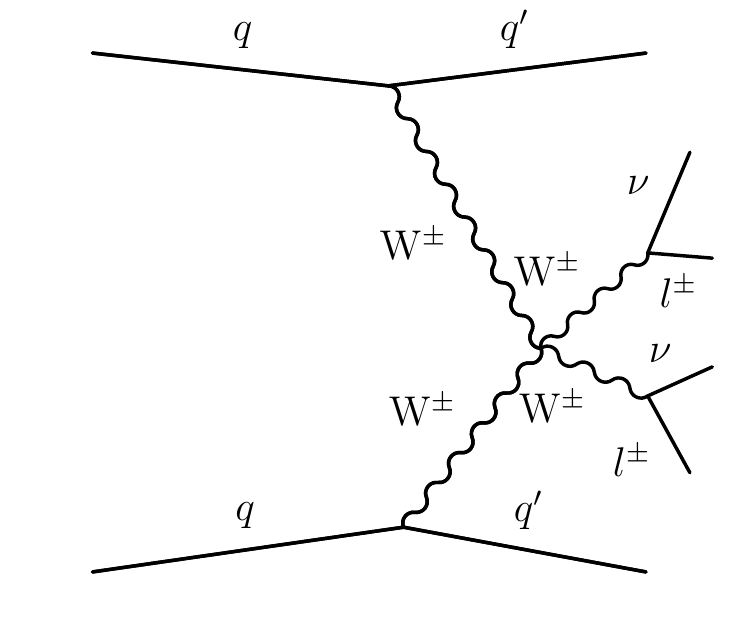}
\includegraphics[width=0.49\textwidth]{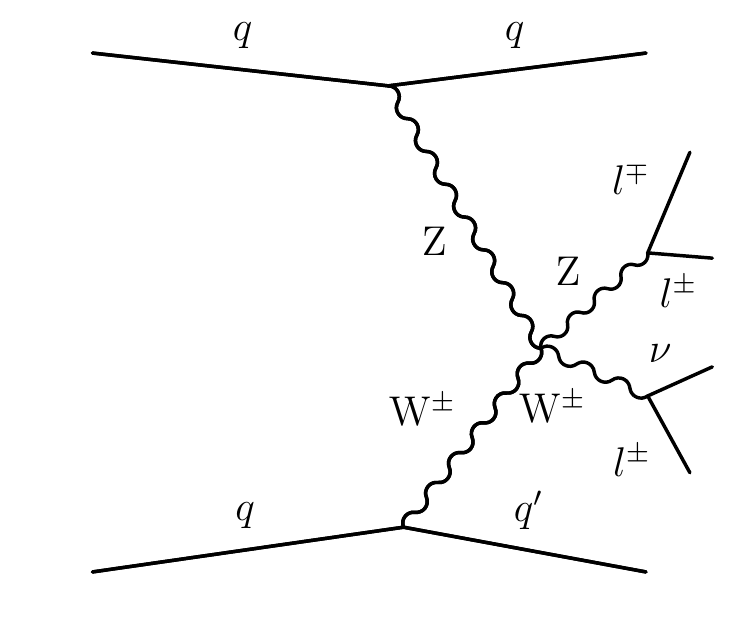}
\caption{Representative Feynman diagrams of a VBS process contributing to the EW-induced production of events containing 
$\WW$ (left) and $\WZ$ (right) boson pairs decaying to leptons, 
and two forward jets.\label{fig:feynman}}
\end{figure*}

\begin{figure*}[htb]
\centering
\includegraphics[width=0.49\textwidth]{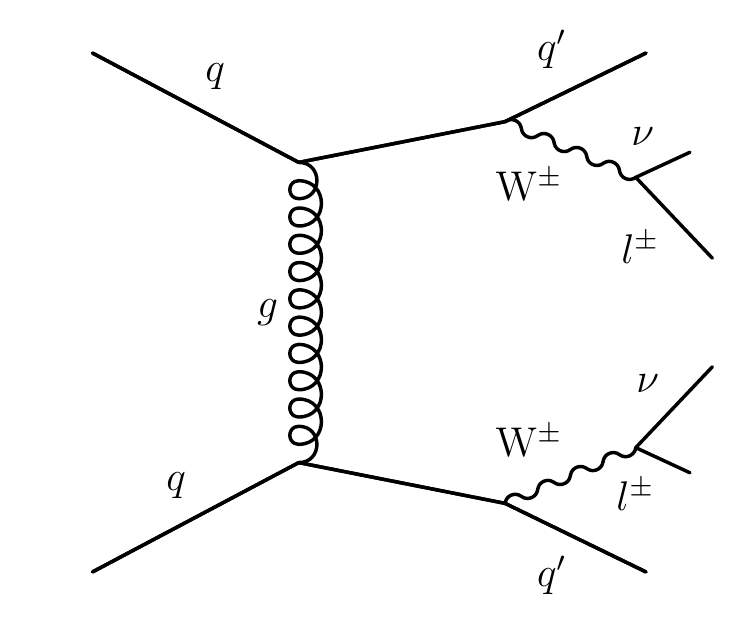}
\includegraphics[width=0.49\textwidth]{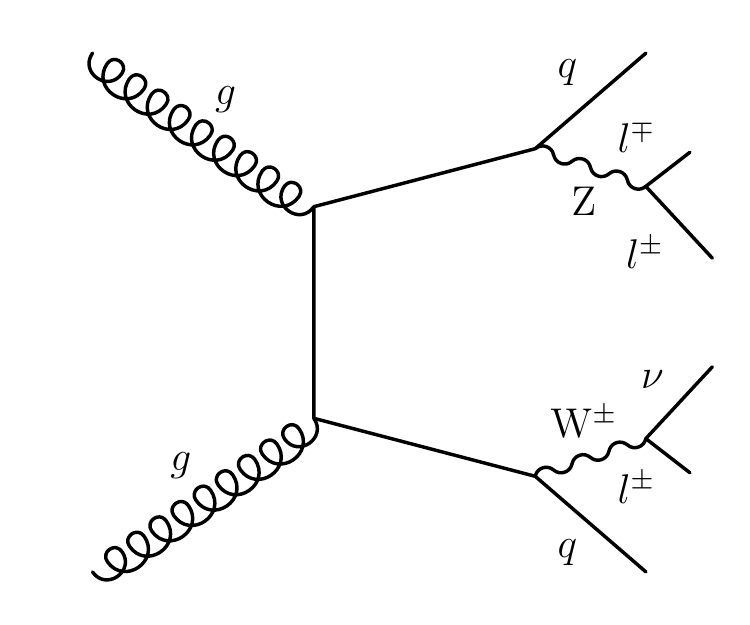}
\caption{Representative Feynman diagrams of the QCD-induced production of $\WW$ (left) and $\WZ$ (right) boson pairs decaying 
to leptons, and two jets.\label{fig:feynman_qcd}}
\end{figure*}

Processes characterized by the presence of two gauge bosons in association with two forward jets are an important background contribution. 
The processes contributing to diboson plus two jets production that proceeds via the EW interaction are 
referred to as EW-induced diboson production, leading to tree-level contributions at $\mathcal{O}(\alpha^4)$, 
where $\alpha$ is the EW coupling. Figure~\ref{fig:feynman} shows representative Feynman diagrams of EW-induced diboson production involving quartic vertices. An additional contribution to the diboson plus two jets production arises via quantum 
chromodynamics (QCD) radiation, leading to tree-level contributions 
at $\mathcal{O}(\alpha^2\alpS^2)$, where $\alpS$ is the strong coupling. This class of processes is referred to 
as QCD-induced diboson production.  Figure~\ref{fig:feynman_qcd} 
shows representative Feynman diagrams of the QCD-induced production. The associated production of a $\PZ$ boson and a single top quark, 
referred to as $\tZq$ production, is also an important background contribution. Additional background 
contributions arise from the $\ttbar$, $\PQt\PW$, $\ttbar\PW$, $\ttbar\PZ$, $\ttbar\gamma$, triple vector 
boson ($\PV\PV\PV$, $\PV=\PW$, $\PZ$), and double parton scattering processes. 

{\tolerance=800 Multiple Monte Carlo (MC) event generators are used to simulate the signal and 
background contributions. The signal and background processes are produced with on-shell particles. Three sets of simulated events for each process are 
needed to match the data taking conditions in the three years. 
The charged Higgs boson signal samples are simulated using 
\MGvATNLO 2.4.2~\cite{Alwall:2014hca,Frederix2012} at leading order (LO) accuracy. 
The predicted signal cross sections are taken at next-to-next-to-LO 
(NNLO) accuracy from the GM model~\cite{Zaro:2002500}.\par}

The SM EW $\WW$ and $\WZ$ processes, where both bosons decay leptonically, are simulated using \MGvATNLO at LO accuracy 
with six EW ($\mathcal{O}(\alpha^6)$) and zero QCD vertices. The same generator is also used to simulate the QCD-induced $\WW$ 
process with four EW and two QCD vertices. Contributions with an initial-state {\cPqb} quark are excluded from the EW $\WZ$ 
simulation because they are considered part of the $\tZq$ background process. Triboson processes, where the $\WZ$ boson pair 
is accompanied by a third vector boson that decays into jets, are included in the EW $\WZ$ simulation. The QCD-induced $\WZ$ 
process is simulated at LO with up to three additional partons in the matrix element calculations using the \MGvATNLO generator 
with at least one QCD vertex at tree level. 
The different jet multiplicities are merged using the MLM scheme~\cite{MLMmerging} 
to match matrix element and parton shower jets, and the inclusive contribution is normalized 
to NNLO predictions~\cite{Grazzini:2016swo}. The interference between the EW and QCD diagrams is also accounted for with \MGvATNLO. 

A complete set of NLO QCD and EW corrections for the leptonic $\WW$ 
scattering process has been computed~\cite{Biedermann:2016yds,Biedermann:2017bss} and they reduce the LO cross section of 
the EW $\WW$ process by 10--15\%, with the 
correction increasing in magnitude with increasing dilepton and dijet 
invariant masses. Similarly, the NLO QCD and EW corrections for the leptonic 
$\WZ$ scattering process have been computed at the orders of $\mathcal{O}(\alpS\alpha^6)$ 
and $\mathcal{O}(\alpha^7)$~\cite{Denner:2019tmn}, reducing the cross sections 
for the EW $\WZ$ process by 10\%. The SM EW $\WW$ and $\WZ$ processes are normalized by applying these $\mathcal{O}(\alpS\alpha^6)$ and 
$\mathcal{O}(\alpha^7)$ corrections to \MGvATNLO LO cross sections.

{\tolerance=800 The $\POWHEG$~v2~\cite{Frixione:2002ik,Nason:2004rx,Frixione:2007vw,Alioli:2008gx,Alioli:2010xd} generator 
is used to simulate the $\ttbar$, $\PQt\PW$, $\PZ\PZ$, and $\PW^{\pm}\PW^{\mp}$ processes at NLO accuracy in QCD. 
Production of $\ttbar\PW$, $\ttbar\PZ$, $\ttbar\gamma$, and $\PV\PV\PV$ events 
is simulated at NLO accuracy in QCD using the $\MGvATNLO$~2.2.2 (2.4.2) generator for the 2016 (2017 and 2018) samples. The $\tZq$ process is simulated in the four-flavor scheme using $\MGvATNLO$ 2.3.3 at next-to-LO (NLO). 
Events in which two hard parton-parton interactions occur within a single $\Pp\Pp$ collision, referred to as double parton scattering $\WW$ production, are generated at LO using $\PYTHIA$ 8.226 (8.230)~\cite{Sjostrand:2014zea} for the 2016 (2017 and 2018) samples.\par}

The NNPDF~2.3 LO \cite{Ball:2012cx} (NNPDF~3.1 NNLO~\cite{Ball:2017nwa}) PDFs are used for generating 2016 
(2017 and 2018) signal samples. The NNPDF~3.0- NLO \cite{Ball:2014uwa} (NNPDF~3.1 NNLO) PDFs are used for 
generating all 2016 (2017 and 2018) background samples. For all processes, the parton showering 
and hadronization are simulated using $\PYTHIA$~8.226 (8.230) for 2016 (2017 and 2018). The modeling 
of the underlying event is done using the CUETP8M1~\cite{Skands:2014pea,Khachatryan:2015pea} 
(CP5~\cite{Sirunyan:2019dfx}) tune for simulated samples corresponding to the 2016 (2017 and 2018) data.

All MC generated events are processed through a simulation of the CMS detector based on 
\GEANTfour~\cite{Geant} and are reconstructed with the same algorithms used for data. The simulated samples include additional interactions in the same and neighboring bunch crossings, referred to as pileup. The additional inelastic events are generated using $\PYTHIA$ with the same underlying event tune as the main interaction and superimposed on the hard-scattering events. The distribution of the number of pileup interactions in the simulation is adjusted 
to match the one observed in the data. 
The average number of interactions per bunch crossing was 23 (32) in 2016 (2017 and 2018) corresponding to an inelastic $\Pp\Pp$ cross-section of 69.2\unit{mb}.

\section{Event reconstruction}
\label{sec:objects}

The primary vertex (PV) is defined as the vertex with the largest value 
of summed physics-object $\pt^2$.  The physics objects are the jets, 
clustered using the jet finding algorithm~\cite{Cacciari:2008gp,Cacciari:2011ma} 
with the tracks assigned to candidate vertices as inputs, and the associated 
missing transverse momentum, taken as the negative vector sum of the $\pt$ of those jets.

The CMS particle-flow (PF) algorithm~\cite{Sirunyan:2017ulk} is used to combine 
the information from the tracker, calorimeters, and muon systems to reconstruct and identify 
charged and neutral hadrons, photons, muons, and electrons (PF candidates). 
The missing transverse momentum vector $\ptvecmiss$ is 
defined as the projection onto the plane perpendicular to the beam axis of the negative 
vector momentum sum of all reconstructed PF candidates in an event. 
Its magnitude is referred to as $\ptmiss$. 

Jets are reconstructed by clustering PF candidates using the anti-\kt 
algorithm~\cite{Cacciari:2008gp} with a distance parameter of 0.4. Additional 
proton-proton interactions within the same or nearby bunch crossings can contribute 
additional tracks and calorimetric energy depositions, increasing the apparent jet 
momentum. To mitigate this effect, tracks identified to be originating from pileup 
vertices are discarded and an offset correction is applied to correct for remaining 
contributions~\cite{Sirunyan:2020foa}. Jet energy corrections are derived from 
simulation studies so that the average measured energy of jets becomes identical 
to that of particle level jets. In situ measurements of the momentum balance in 
dijet, photon+jet, Z+jet, and multijet events are used to determine any residual 
differences between the jet energy scale in data and in simulation, and appropriate 
corrections are made~\cite{Khachatryan:2016kdb}. Corrections to jet energies to 
account for the detector response are propagated to $\ptmiss$~\cite{Sirunyan:2019kia}. 
Jets with transverse momentum $\pt>30\GeV$ and $\abs{\eta}<4.7$ are included in the analysis. 

Events with at least one jet with $\pt>20\GeV$ and $\abs{\eta}<2.4$ that 
is consistent with the fragmentation of a bottom quark are rejected to reduce the number of top quark background events. 
The \textsc{DeepCSV} \PQb  tagging algorithm~\cite{Sirunyan:2017ezt} is used for this selection. 
For the chosen working point, the efficiency of the algorithm to select \PQb  quark jets is about 72\% and the rate for incorrectly tagging jets 
originating from the hadronization of gluons or $\cPqu$, $\cPqd$, $\cPqs$ quarks 
is about 1\%. The rate for incorrectly tagging jets originating from the hadronization of $\cPqc$ quarks is about 10\%.

Events with at least one reconstructed hadronic decay of a $\tau$ lepton, denoted as $\tauh$, with $\pt>18\GeV$ and $\abs{\eta}<2.3$, 
are rejected to reduce the contribution of diboson processes with $\tauh$ decays. The $\tauh$ decays are reconstructed using the hadrons-plus-strips algorithm~\cite{Khachatryan:2015dfa}. 

Electrons and muons are reconstructed by associating a track reconstructed 
in the tracking detectors with either a cluster of energy deposits in the 
ECAL~\cite{Khachatryan:2015hwa, Sirunyan:2020ycc} or a track in the muon system~\cite{Sirunyan:2018fpa}. 
Electrons (muons) must pass loose identification criteria with $\pt>10\GeV$ and $\abs{\eta}<2.5$ (2.4) to be selected for the analysis. 
At the final stage of the lepton selection, tight working points, following the definitions provided in Refs.~\cite{Khachatryan:2015hwa,Sirunyan:2020ycc,Sirunyan:2018fpa}, 
are chosen for the identification criteria, including requirements on the impact parameter 
of the candidates with respect to the PV and their isolation with 
respect to other particles in the event~\cite{Sirunyan:2018egh}. 
For electrons, the background contribution arising from charge misidentification is not negligible. 
The sign mismeasurement is evaluated using three observables that measure the electron curvature applying different methods as discussed in Ref.~\cite{Khachatryan:2015hwa}. Requiring all three charge evaluations to agree reduces this background contribution 
by a factor of four (six) with an efficiency of about 97 (90)\% in the barrel (endcap) region. 
The sign mismeasurement is negligible for muons~\cite{Chatrchyan:2009ae,Sirunyan:2019yvv}.  

\section{Event selection}
\label{sec:selection}

Collision events are collected using single-electron and single-muon
triggers that require the presence of an isolated lepton with $\pt>27$ and 24\GeV, respectively~\cite{Sirunyan:2021zrd}. In addition, a set of dilepton triggers with lower $\pt$ thresholds 
is used, ensuring a trigger efficiency above 99\% for events that satisfy the 
subsequent offline selection~\cite{Sirunyan:2021zrd}.

Several selection requirements are used to isolate the $\WW$ and $\WZ$ topologies defining the signal 
regions (SRs), while reducing the contributions from background processes~\cite{Sirunyan:2020gyx}. 
Candidate events must contain exactly two isolated same-sign charged leptons or exactly three isolated charged 
leptons with $\pt>10\GeV$, and at least two jets with $\abs{\eta}<4.7$ and 
the leading jet $\pt^{\mathrm{j}}>50\GeV$. To exclude the selected electrons and muons from the jet sample, the
jets are required to be separated from the identified leptons by $\Delta R = \sqrt{\smash[b]{(\Delta \eta)^{2} + (\Delta \phi)^{2}}} > 0.4$, where $\phi$ is the azimuthal angle in radians.

For the $\WZ$ candidate events, one of the oppositely charged same-flavor leptons from the $\PZ$ boson 
candidate is required to have $\pt>25\GeV$ and the other $\pt>10\GeV$ with the 
invariant mass of the dilepton pair $\mll$ satisfying $\abs{\mll - m_{\PZ}}<15\GeV$. 
For candidate events with three same-flavor leptons, the oppositely charged lepton pair with the 
invariant mass closest to the world-average $\PZ$ boson mass $m_{\PZ}$~\cite{10.1093/ptep/ptaa104} is selected 
as the $\PZ$ boson candidate. The third lepton associated with the $\PW$ boson is required to have $\pt>20\GeV$. 
In addition, the trilepton invariant mass $\mlll$ is required to exceed 100\GeV to exclude a region where production of $\PZ$ bosons with final-state photon radiation is expected to contribute.   

One of the leptons in the same-sign $\WW$ candidate events is required to have 
$\pt>25\GeV$ and the other $\pt>20\GeV$. The value of $\mll$ must be greater than 20\GeV. Candidate events in the dielectron final state 
with $\abs{\mll-m_{\PZ}}<15\GeV$ are rejected to reduce the number of $\PZ$ boson background events where the sign of one of the electron candidates is 
misidentified. 

The VBF topology is targeted by requiring the two highest $\pt$ jets to have 
a mass $\mjj>500\GeV$ and a pseudorapidity separation $\detajj>2.5$. 
The $\PW$ and $\PZ$ bosons in the VBF topologies are mostly produced in the central rapidity 
region with respect to the two selected jets. The candidate $\WW$ ($\WZ$) events are 
required to satisfy $\mathrm{max}(\mathrm{z}_{\ell}^{*})<0.75 (1.0)$, where 
$\mathrm{z}_{\ell}^{*}=\abs{\eta^{\ell} - (\eta^{\jet_{1}} + \eta^{\jet_{2}})/2}/\detajj$ 
is the Zeppenfeld variable~\cite{Rainwater:1996ud} for one of the selected leptons. Here $\eta^{\ell}$ is the 
pseudorapidity of the lepton, and $\eta^{\jet_{1}}$ and 
$\eta^{\jet_{2}}$ are the pseudorapidities of the two candidates VBF jets. 

The $\ptmiss$ is required to exceed 30\GeV for both SRs. The selection requirements used to define the same-sign $\WW$ and 
$\WZ$ SRs are summarized in Table~\ref{tab:selectioncutsSR}. 

\begin{table*}[htbp]
\centering
  \topcaption{Summary of the selection requirements defining the $\WW$ and $\WZ$ SRs. The 
  looser lepton $\pt$ requirement in the $\WZ$ selection refers to the trailing lepton from the $\PZ$ boson decays. 
  The $\abs{\mll - m_{\PZ}}$ requirement is applied only to the dielectron final state in the $\WW$ SR.\label{tab:selectioncutsSR}}
\begin{tabular} {lcc}
\hline
  Variable & $\WW$ & $\WZ$ \\
\hline
Leptons                 & 2 leptons, $\pt>25/20\GeV$ & 3 leptons, $\pt>25/10/20\GeV$ \\
$\pt^\jet$              & $>$50/30\GeV               & $>$50/30\GeV                  \\
$\abs{\mll - m_{\PZ}}$ & $>$15\GeV ($\Pe\Pe$)       & $<$15\GeV		     \\
$\mll$                  & $>$20\GeV                  & \NA			     \\
$\mlll$                 & \NA                        & $>$100\GeV		     \\
$\ptmiss$               & $>$30\GeV		     & $>$30\GeV		     \\
\PQb  jet veto        & Required		     & Required                      \\
$\tauh$  veto           & Required		     & Required 		     \\
$\zepmax$               & $<$0.75		     & $<$1.0  		             \\
$\mjj$                  & $>$500\GeV                 & $>$500\GeV		     \\
$\detajj$               & $>$2.5		     & $>$2.5  		             \\
\hline
\end{tabular}
\end{table*}

\section{Background estimation}
\label{sec:backgrounds}

A combination of methods based on simulation and on control regions (CRs) in data is used to 
estimate background contributions. By inverting some of the requirements in Table~\ref{tab:selectioncutsSR} we select background-enriched CRs.  
Uncertainties related to the theoretical and experimental predictions are 
estimated as described in Section~\ref{sec:systematics}. 

The nonprompt lepton backgrounds originating from leptonic decays of heavy quarks, 
hadrons misidentified as leptons, and electrons from photon conversions are 
suppressed by the identification and isolation requirements imposed on leptons. The remaining contribution from the nonprompt lepton background is dominant in the $\WW$ SR and is estimated directly from data following the technique described in Ref.~\cite{Khachatryan:2014sta}, using events selected by the final selection criteria, except for one of the leptons, which is requested to pass a looser criterion having failed the nominal selection. The yield in this sample is extrapolated to the signal region 
using the efficiencies for such loosely identified leptons to pass the standard 
lepton selection criteria. This efficiency is calculated in a sample of events 
dominated by dijet production. An uncertainty of 20\% is assigned for the nonprompt lepton 
background normalization to include possible differences in the composition of jets 
between the data sample used to derive these efficiencies and the data samples 
in the $\WW$ and $\WZ$ SRs~\cite{Sirunyan:2018egh}.

The background contribution from the electron sign mismeasurement is estimated from the simulation by applying 
a data-to-simulation efficiency correction due to electrons with sign mismeasurement. These corrections are determined using $\PZ \to \Pe\Pe$ events in the $\PZ$ boson peak region that were recorded with 
independent triggers. These corrections amount to 40\% for data collected in 2017 and 2018, while they are negligible for 2016 data. The electron sign mismeasurement rate is about 0.01 (0.3)\% in the barrel (endcap) region~\cite{Khachatryan:2015hwa, Sirunyan:2020ycc}.

Three CRs are used to select nonprompt lepton, $\tZq$, and $\PZ\PZ$ 
background-enriched events to further estimate the normalization of these background processes from data. The nonprompt lepton CR is defined by requiring the same selection as for the 
$\WW$ SR, but with the \PQb jet veto requirement inverted. The selected events are enriched in the nonprompt lepton background coming mostly from semileptonic $\ttbar$ events.
Similarly, the $\tZq$ CR is defined by requiring the same selection as for the 
$\WZ$ SR, but with the \PQb jet veto requirement inverted. The selected events 
are dominated by the $\tZq$ background process. 
Finally, the $\PZ\PZ$ CR selects events with two opposite-sign same-flavor lepton pairs with the same VBS-like requirements. 
The three CRs are used together with the SRs to constrain the normalization of the nonprompt lepton, $\tZq$, and $\PZ\PZ$ 
background processes from data. All other background processes are estimated from simulation after 
applying corrections to account for the small differences between data and simulation. 
The shapes of the $\tZq$ and $\PZ\PZ$ background processes are estimated from simulation as well. 

\begin{figure}[htbp]
\begin{center}
\includegraphics[width=0.49\textwidth]{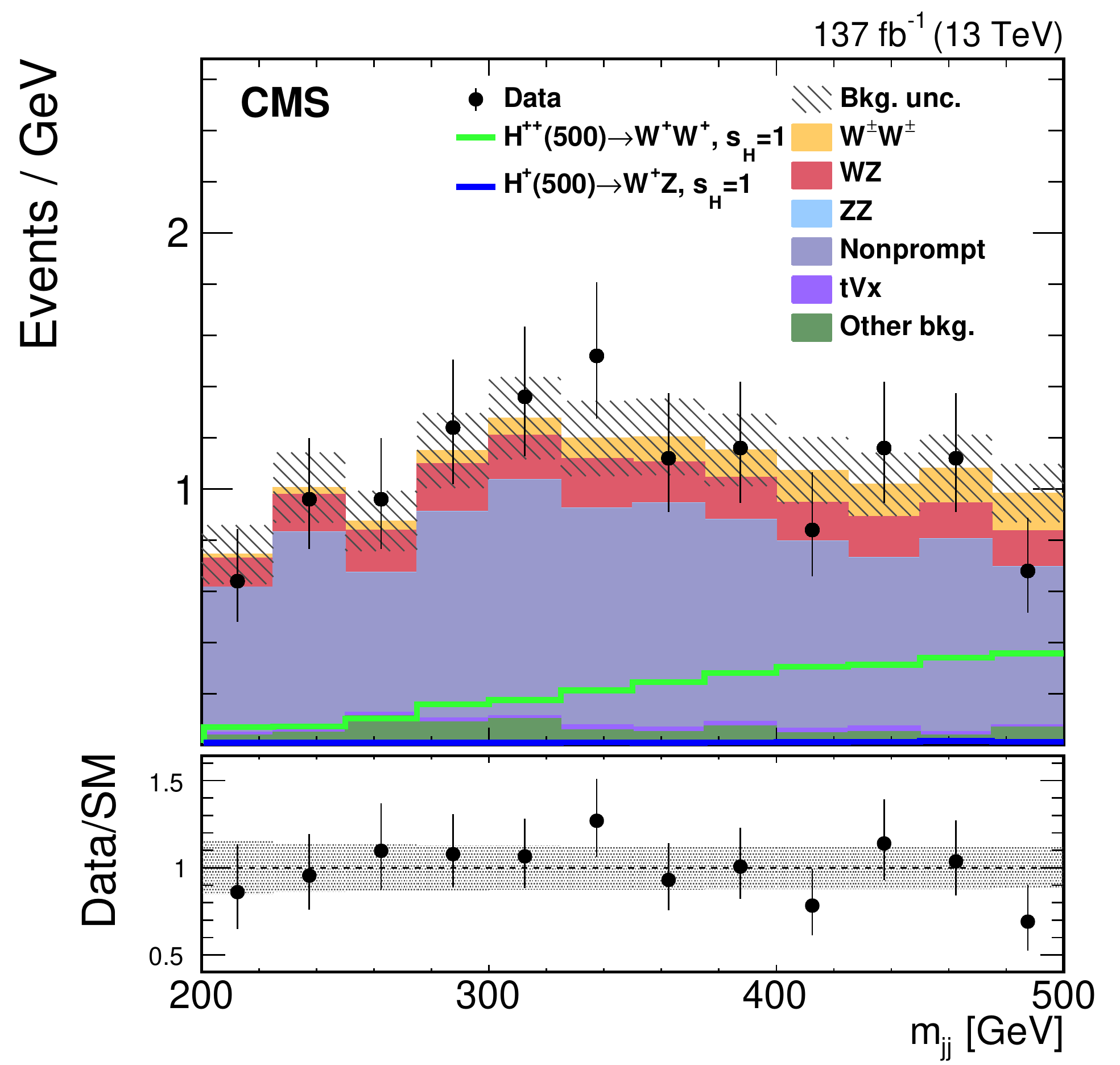}
\includegraphics[width=0.49\textwidth]{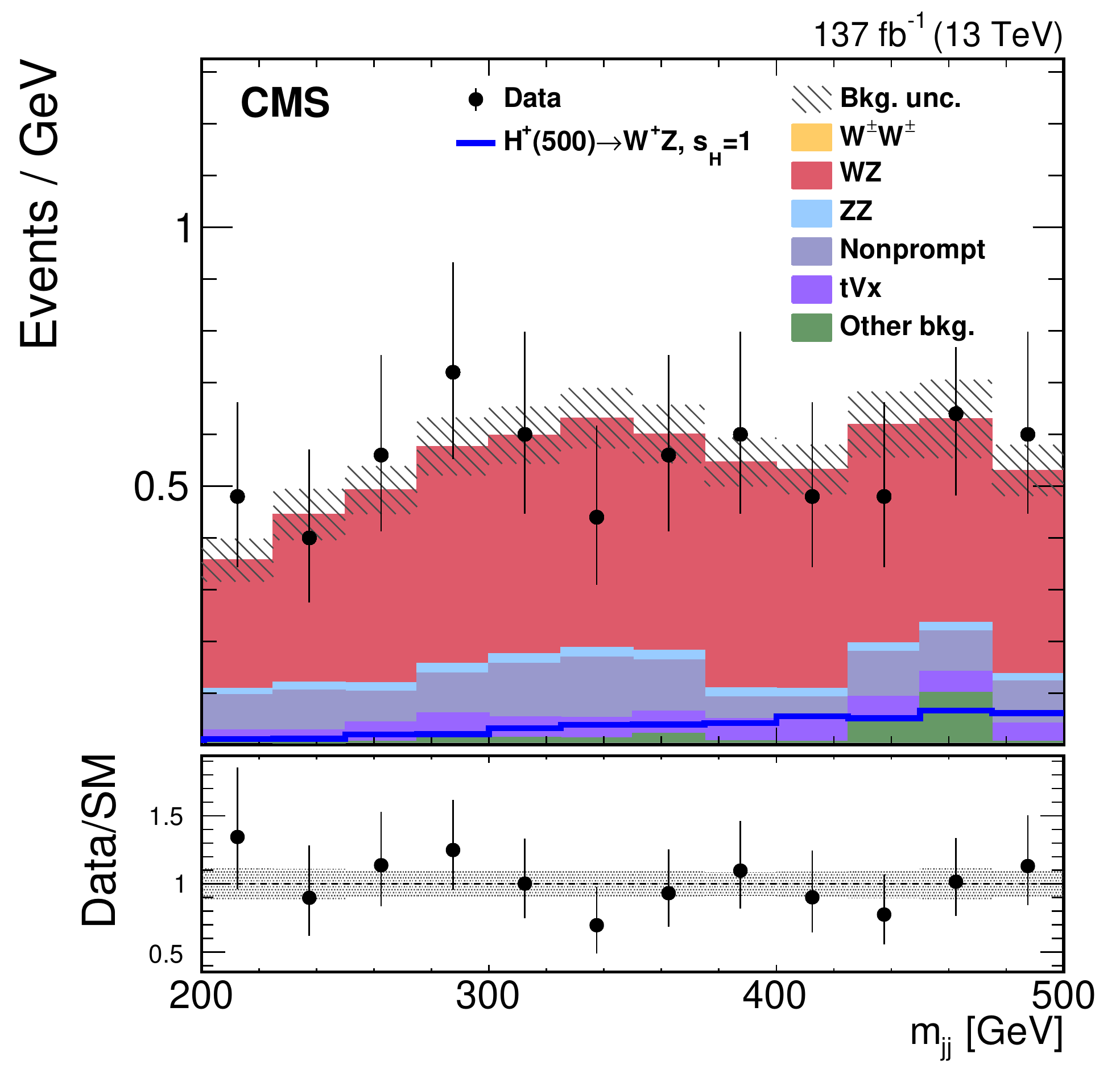}
\caption{The $\mjj$ distributions after requiring the same selection as for the $\PW\PW$ (\cmsLeft) and $\PW\PZ$ (\cmsRight) SRs, 
but with a requirement of $200<\mjj<500\GeV$. 
The predicted yields are shown with their best fit normalizations from the simultaneous fit (described in Section~\ref{sec:signal}) for the background-only hypothesis \ie, assuming no contributions from the  $\PHpm$ and $\PHpmpm$ processes. 
Vertical bars on data points represent the statistical uncertainty in the data. 
The histograms for $\tVx$ backgrounds include the contributions from $\ttbar\PV$ and $\tZq$ processes. 
The histograms for other backgrounds  include the contributions from double parton scattering, $\PV\PV\PV$, 
and from oppositely charged dilepton final states from $\ttbar$, $\PQt\PW$, $\PW^{+}\PW^{-}$, and Drell--Yan processes. 
The overflow is included in the last bin. The lower panels show the ratio of 
the number of events observed in data to that of the total SM prediction. 
The hatched gray bands represent the uncertainties in the predicted yields. The solid lines show 
the signal predictions for values of $s_{\PH}=1.0$ and $m_{\PH_{5}}=500\GeV$ in the GM model.
\label{fig:ssww_qcd}}
\end{center}
\end{figure}

The prediction for the QCD $\WZ$ background process is validated in a CR defined by 
requiring the same selection as for the $\WZ$ SR, but with a requirement of $200<\mjj<500\GeV$. The predicted yields are shown with their best fit normalizations from the simultaneous fit (described in Section~\ref{sec:signal}) for the background-only hypothesis \ie, assuming no contributions from the  $\PHpm$ and $\PHpmpm$ processes. 
Good agreement between the data and post-fit predicted yields is observed in this CR as can be seen in Fig.~\ref{fig:ssww_qcd}. 

\section{Signal extraction}
\label{sec:signal}
A binned maximum-likelihood fit is performed using the $\WW$ and $\WZ$ SRs, and the nonprompt lepton, $\tZq$, and $\PZ\PZ$ CRs to discriminate between the signal and the remaining backgrounds. Signal contributions with electrons and muons produced in the decay of a $\tau$ lepton are included. The normalization factors for the 
$\tZq$ and $\PZ\PZ$ background processes, affecting both the SRs and CRs, are included as free parameters in the 
maximum-likelihood fit together with the signal strength. The SM $\WW$ ($\WZ$) 
contribution is obtained from the sum of the EW $\WW$ ($\WZ$), QCD $\WW$ 
($\WZ$), and the interference contributions according to the SM predictions~\cite{Sirunyan:2020gyx} and allowed to vary within the uncertainties. 

{\tolerance=800 The diboson transverse mass ($\mT^{\PV\PV}$) is constructed from the four-momentum of the selected charged leptons and the $\ptvecmiss$. The four-momentum of the neutrino system is defined using the $\ptvecmiss$, assuming that the values of the longitudinal component of the momentum and the mass are zero. The value of $\mT^{\PV\PV}$, defined as

\begin{equation}
\mT^{\PV\PV} = \sqrt{{{\biggl(\sum\nolimits_i E_{i}\biggr)^2-\biggl(\sum\nolimits_i p_{z,i}\biggr)^2}}},
\label{eq:mtvv}
\end{equation}

where $E_{i}$ and $p_{z,i}$ are the energies and longitudinal components of the momenta of the leptons 
and neutrino system from the decay of the gauge bosons in the event, is effective in discriminating between 
the resonant signal and nonresonant background processes. The value of $\mjj$ is effective in discriminating between 
all non-VBS processes and the signal (plus EW $\PV\PV$) processes because VBF and VBS 
topologies typically exhibit large values for the dijet mass. A two-dimensional distribution is used 
in the fit for the $\WW$ SR with 8 bins in $\mT^{\PV\PV}$ ([0, 250, 350, 450, 550, 650, 850, 1050, $\infty$]\GeV) 
and 4 bins in $\mjj$ ([500, 800, 1200, 1800, $\infty$]\GeV). Similarly, a two-dimensional distribution is used 
in the fit for the $\WZ$ SR with 7 bins in $\mT^{\PV\PV}$ ([0, 325, 450, 550, 650, 850, 1350, $\infty$]\GeV) 
and 2 bins in $\mjj$ ([500, 1500, $\infty$]\GeV). The $\mjj$ distribution is used for the CRs in the fit 
with 4 bins ([500, 800, 1200, 1800, $\infty$]\GeV).\par}

A profile likelihood technique is used where systematic uncertainties are represented by nuisance parameters~\cite{Cowan:2010js}. 
For each individual bin, a Poisson likelihood term describes the fluctuation of the data around
the expected central value, which is given by the sum of the contributions from signal and 
background processes. The systematic uncertainties 
are treated as nuisance parameters and are profiled with the shape and 
normalization of each distribution varying within the respective uncertainties in the fit. The 
normalization uncertainties are treated as log-normal nuisance parameters. Correlation across 
bins is taken into account. The uncertainties affecting the shapes of the distributions are 
modeled in the fit as nuisance parameters with external Gaussian constraints. 
The dominant nuisance parameters are not significantly constrained by the data, \ie, 
the normalized nuisance parameter uncertainties are close to unity.

\section{Systematic uncertainties}
\label{sec:systematics}

Several sources of systematic uncertainty are taken into account in the signal extraction procedure. 
For each source of uncertainty,  the effects on the 
signal and background distributions are considered to be correlated.

The total Run~2 (2016--2018) integrated luminosity has an uncertainty of 1.8\%, the improvement in precision relative to Refs.~\cite{CMS-PAS-LUM-17-001,CMS-PAS-LUM-17-004,CMS-PAS-LUM-18-002} reflecting the (uncorrelated) time evolution of some systematic effects.

The simulation of pileup events assumes an inelastic $\Pp\Pp$ cross section of 69.2\unit{mb}, 
with an associated uncertainty of 5\%~\cite{Sirunyan:2018nqx}, which has 
an impact on the expected signal and background yields of about 1\%.

Discrepancies in the lepton reconstruction and identification
efficiencies between data and simulation are corrected by applying 
scale factors to all simulation samples.
These scale factors, which depend on the $\pt$ and $\eta$ for both electrons and muons, are determined using $\PZ \to \ell\ell$ 
events in the $\PZ$ boson peak region that were recorded with 
independent triggers~\cite{Sirunyan:2018fpa,Khachatryan:2015hwa,Sirunyan:2019bzr}. 
The uncertainty in the determination of the trigger efficiency leads to an uncertainty smaller than 1\% in the expected signal yield. The trigger efficiency in the simulation is corrected to account for the effect of a gradual time shifts in the forward region in the ECAL endcaps for the 2016 and 2017 data~\cite{Sirunyan:2020zal}. The uncertainty in this correction is included in the trigger efficiency uncertainty.
The lepton momentum scale uncertainty is computed by varying the lepton momenta 
in simulation with their uncertainties, and repeating the 
analysis selection. The resulting uncertainties in the yields are $\approx$1\% for both electrons and muons. 
These uncertainties are assumed to be correlated across the three data sets.

The uncertainty in the calibration of the jet energy scale (JES) directly 
affects the acceptance of the jet multiplicity requirement and the $\ptmiss$ 
measurement. These effects are estimated by shifting the JES in the simulated samples 
up and down by one standard deviation. 
The uncertainty in the jet energy resolution (JER) smearing applied to 
simulated samples to match the $\pt$ resolution measured in data causes both 
a change in the normalization and in the shape of the distributions. 
The overall uncertainty in the JES and JER is 2--5\%, depending on $\pt$ and 
$\eta$~\cite{Khachatryan:2016kdb,CMS-DP-2020-019}, and its impact on the 
expected signal and background yields is about 3\%. 

The \cPqb\ tagging efficiency in the simulation is corrected using scale factors determined 
from data~\cite{Sirunyan:2017ezt}. These values are estimated separately for 
correctly and incorrectly tagged jets. Each set of values results in uncertainties in the \cPqb\ tagging 
efficiency of about 1--4\% depending on $\pt$ and $\eta$, and the impact on the expected signal and background yields is about 1\%. 
The uncertainties in the JER, JES and \cPqb\ tagging are treated as 
uncorrelated across the three data taking years, since the detector conditions 
have changed among the three years.

The theoretical uncertainties associated with the choice of the renormalization and factorization scales are estimated 
by varying these scales independently up and down by a factor of 
two from their nominal values. The envelope of the resulting distributions, excluding the two extreme variations  where one scale is varied up and the other one down, is taken as the uncertainty~\cite{Catani:2003zt,Cacciari:2003fi}. 
The variations of the PDF set and $\alpS$ are used to estimate 
the corresponding uncertainties in the yields of the signal and background processes, 
following Refs.~\cite{Ball:2014uwa,Butterworth:2015oua}. The uncertainty in the yields due to missing higher-order 
EW corrections in the GM model is estimated to be 7\%~\cite{Zaro:2002500}. These theoretical uncertainties may affect both the estimated signal and background rates. The statistical uncertainties that are associated with the limited number of 
simulated events and data events used to estimate the nonprompt lepton background are also considered 
as systematic uncertainties.

A summary of the impact of the systematic uncertainties on the signal strength, $\mu$, defined as 
the ratio of the observed charged Higgs signal yield to the expected yield, is shown in 
Table~\ref{tab:impactsssww_comb_fiducial6_mH500} for the case of a background-only simulated data set, \ie, 
assuming no contributions from the  $\PHpm$ and $\PHpmpm$ processes. Table~\ref{tab:impactsssww_comb_fiducial6_mH500} also shows systematic uncertainties including a charged Higgs 
boson signal for values of $s_{\PH}=1.0$ and $m_{\PH_{5}}=500\GeV$ in the GM model. The impacts shown in Table~\ref{tab:impactsssww_comb_fiducial6_mH500} result from a fit to two simulated 
samples: background-only (first column, expected $\mu = 0$) and signal-plus-background (second column, expected $\mu = 1$). They differ from the 
impacts in percent on the expected signal and background yields given 
above, which are  estimated before the fit.
The total systematic uncertainty is smaller for the background-only simulated data set 
because the uncertainties partially cancel out between the SRs and the CRs for the background processes. 

\begin{table*}[htbp]
\centering
\caption{
Summary of the impact of the systematic uncertainties on the extracted signal strength; for the case of a background-only simulated data set, \ie, assuming no contributions from the 
$\PHpm$ and $\PHpmpm$ processes, and including a charged Higgs boson signal for values of $s_{\PH}=1.0$ and $m_{\PH_{5}}=500\GeV$ in the GM model. The impacts shown result from a fit to two simulated 
samples: background-only (first column, expected $\mu = 0$) and signal-plus-background (second column, expected $\mu = 1$).
\label{tab:impactsssww_comb_fiducial6_mH500}} 
{
\begin{tabular}{lcc}
\hline
\multirow{2}{*}{Source of uncertainty}       & $\Delta \mu$ & $\Delta \mu$ \\
                                             & background-only   &  $s_{\PH}=1.0$ and $m_{\PH_{5}}=500\GeV$     \\
\hline
Integrated luminosity           &  0.002 &  0.019 \\
Pileup  		        &  0.001 &  0.001 \\
Lepton measurement              &  0.003 &  0.033 \\
Trigger                         &  0.001 &  0.007 \\
JES and JER                     &  0.003 &  0.006 \\
\cPqb tagging   	        &  0.001 &  0.006 \\
Nonprompt rate                  &  0.002 &  0.002 \\
$\WW/\WZ$ rate                  &  0.014 &  0.015 \\
Other prompt background rate    &  0.002 &  0.015 \\
Signal rate                     &   \NA  &  0.064 \\
Limited sample size                     &  0.005 &  0.005 \\[\cmsTabSkip]

Total systematic uncertainty    &  0.016 &  0.078 \\
Statistical uncertainty         &  0.021 &  0.044 \\[\cmsTabSkip]

Total uncertainty               &  0.027 &  0.090 \\
\hline
\end{tabular}
}
\end{table*}

\section{Results}
\label{sec:results}

\begin{figure*}[htbp]
\begin{center}
\includegraphics[width=0.49\textwidth]{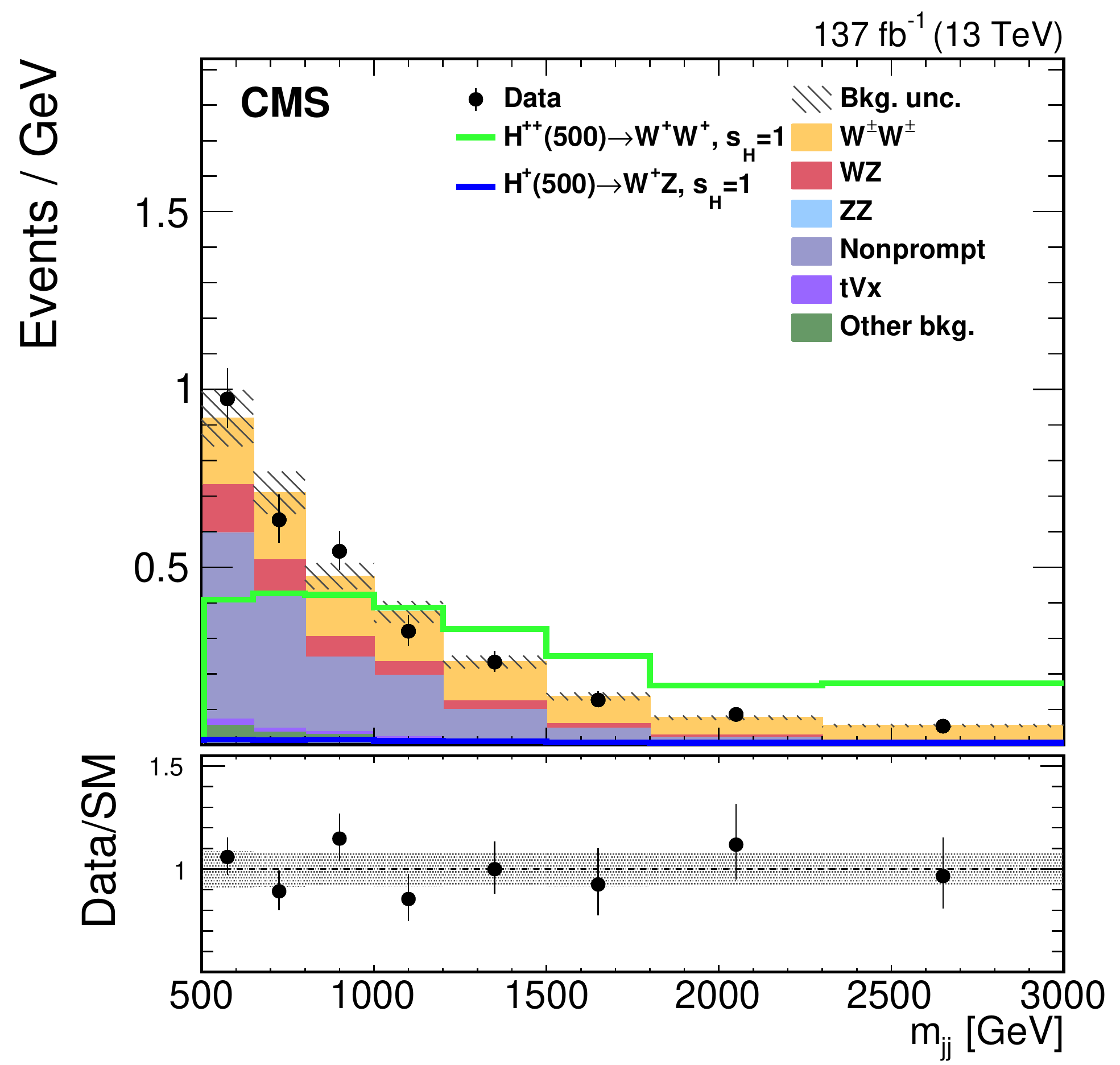}
\includegraphics[width=0.49\textwidth]{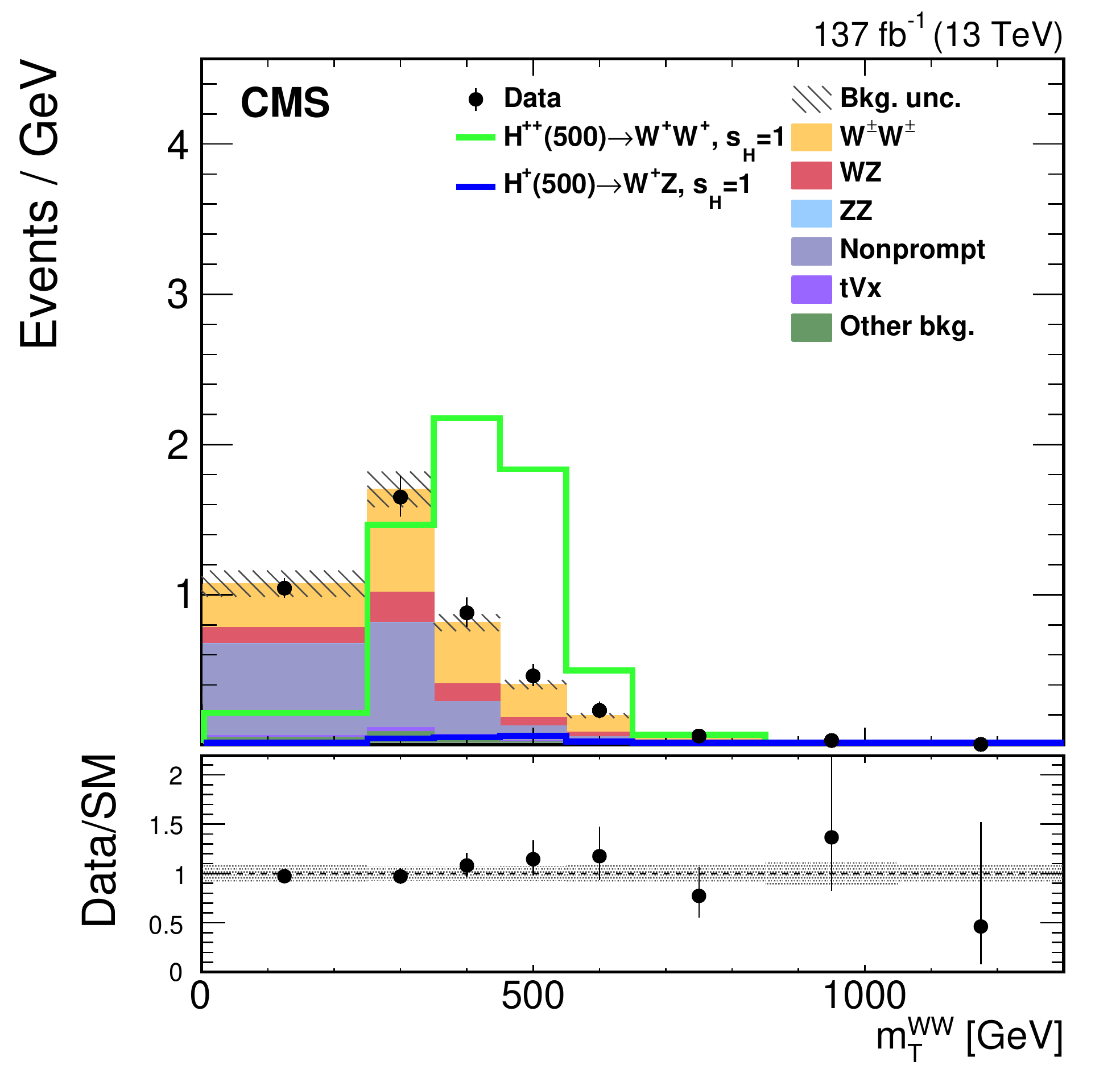}
\includegraphics[width=0.49\textwidth]{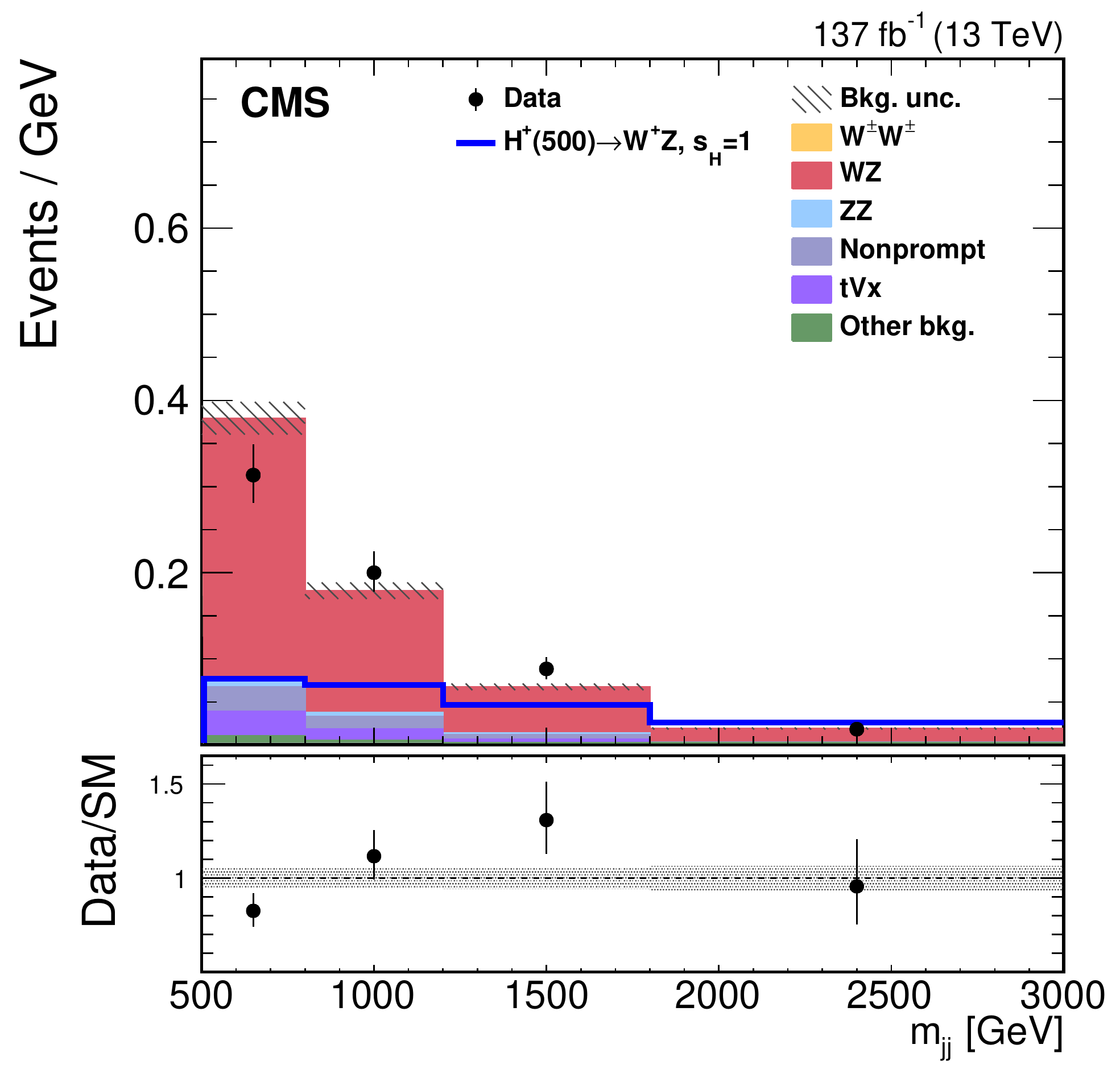}
\includegraphics[width=0.49\textwidth]{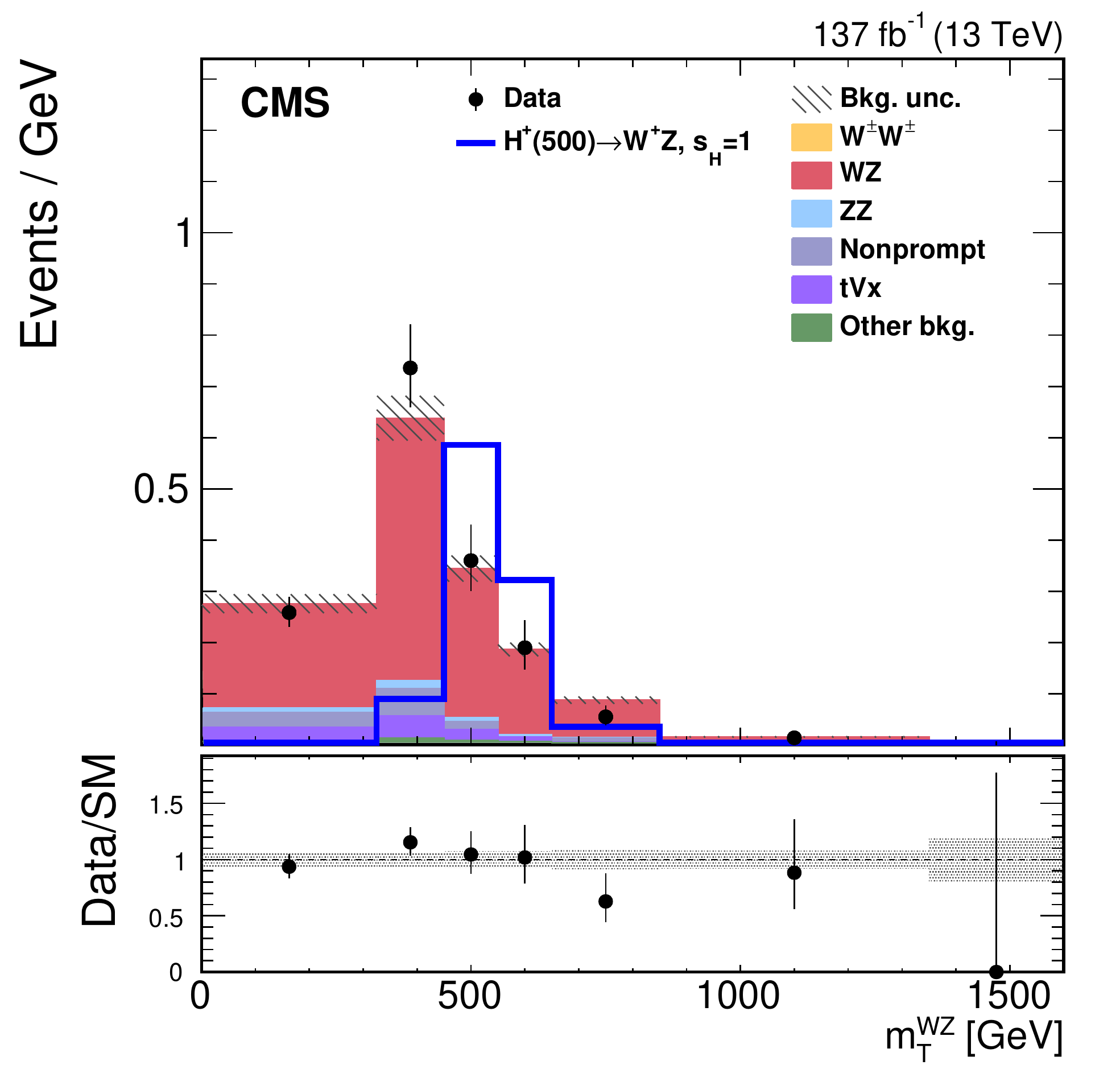}
\caption{
The $\mjj$ (upper left) and $\mT^{\PW\PW}$ (upper right) distributions in the $\PW\PW$ SR, and 
the $\mjj$ (lower left) and $\mT^{\PW\PZ}$ (lower right) distributions in the $\PW\PZ$ SR for signal, backgrounds, and data. 
The predicted yields are shown with their best fit normalizations from the simultaneous fit for the background-only hypothesis, \ie, assuming no contributions from the  $\PHpm$ and $\PHpmpm$ processes. 
Vertical bars on data points represent the statistical uncertainty in the data. 
The histograms for $\tVx$ backgrounds include the contributions from $\ttbar\PV$ and $\tZq$ processes. 
The histograms for other backgrounds  include the contributions from double parton scattering, $\PV\PV\PV$, 
and from oppositely charged dilepton final states from $\ttbar$, $\PQt\PW$, $\PW^{+}\PW^{-}$, and Drell--Yan processes. 
The overflow is included in the last bin. The lower panels show the ratio of 
the number of events observed in data to that of the total SM prediction. 
The hatched gray bands represent the uncertainties in the predicted yields. The solid lines show 
the signal predictions for values of $s_{\PH}=1.0$ and $m_{\PH_{5}}=500\GeV$ in the GM model.
    \label{fig:ssww_mtvv}}
\end{center}
\end{figure*}

\begin{figure*}[htbp]
\begin{center}
\includegraphics[width=0.99\textwidth]{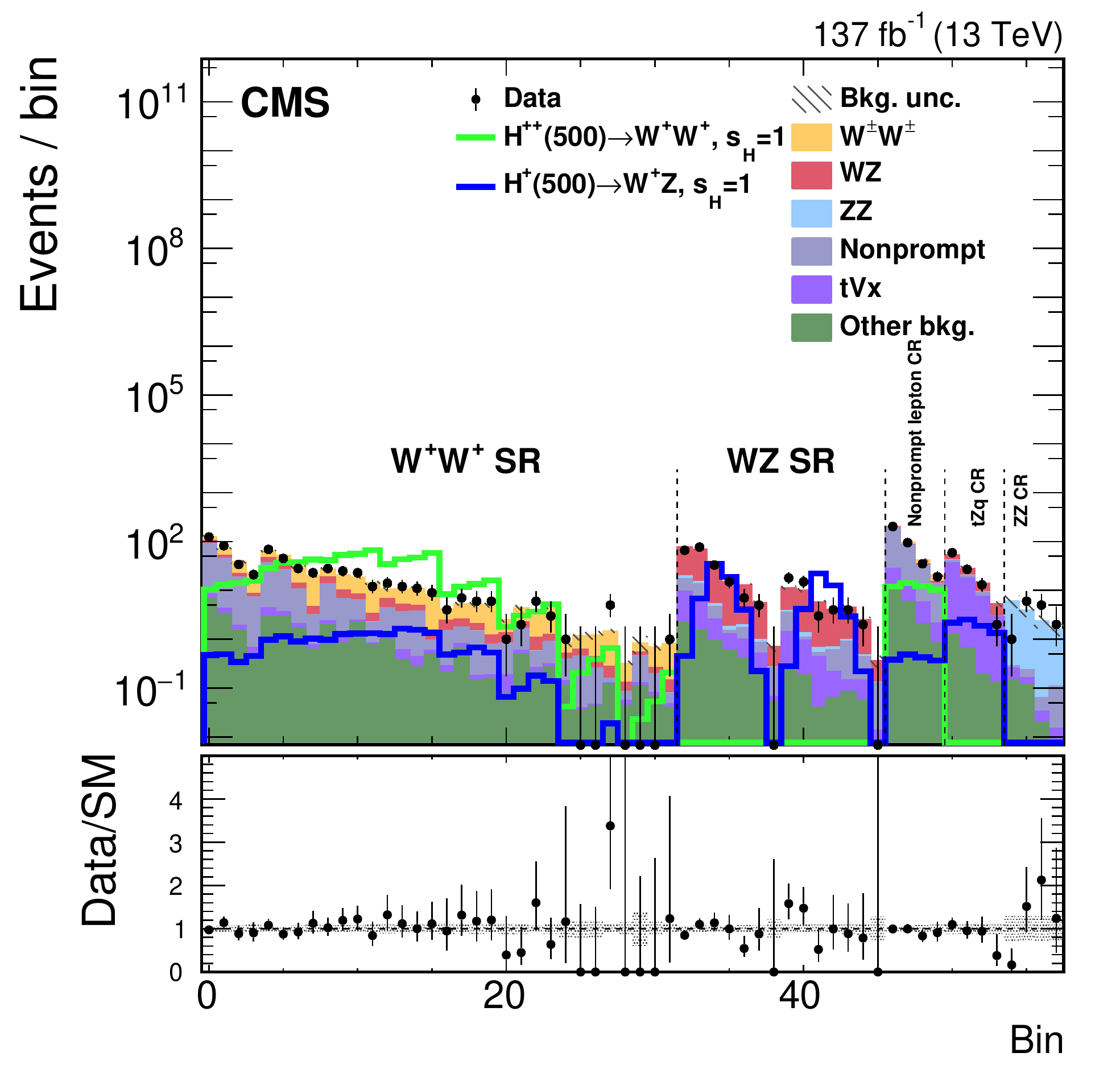}
\caption{Distributions for signal, backgrounds, and data for the bins used in the simultaneous fit. 
The bins 1--32 (4$\times$8) show the events in the $\PW\PW$ SR ($\mjj \times \mT$), the bins 33--46 (2$\times$7) show the events in the $\PW\PZ$ SR ($\mjj \times \mT$), 
the 4 bins 47--50 show the events in the nonprompt lepton CR ($\mjj$), 
the 4 bins 51--54 show the events in the $\tZq$ CR ($\mjj$), 
and the 4 bins 55--58 show the events in the $\PZ\PZ$ CR ($\mjj$). The predicted yields are shown with their best fit normalizations from the simultaneous fit  for the background-only hypothesis, \ie, assuming no contributions from the  $\PHpm$ and $\PHpmpm$ processes. 
Vertical bars on data points represent the statistical uncertainty in the data. 
The histograms for $\tVx$ backgrounds include the contributions from $\ttbar\PV$ and $\tZq$ processes. 
The histograms for other backgrounds  include the contributions from double parton scattering, $\PV\PV\PV$, 
and from oppositely charged dilepton final states from $\ttbar$, $\PQt\PW$, $\PW^{+}\PW^{-}$, and Drell--Yan processes. 
The overflow is included in the last bin in each corresponding region. The lower panels show the ratio of 
the number of events observed in data to that of the total SM prediction. 
The hatched gray bands represent the uncertainties in the predicted yields. 
The solid lines show the signal predictions for values of $s_{\PH}=1.0$ and $m_{\PH_{5}}=500\GeV$ in the GM model.
    \label{fig:ssww_datacard}}
\end{center}
\end{figure*}

\begin{figure}[htbp]
\begin{center}
\includegraphics[width=0.49\textwidth]{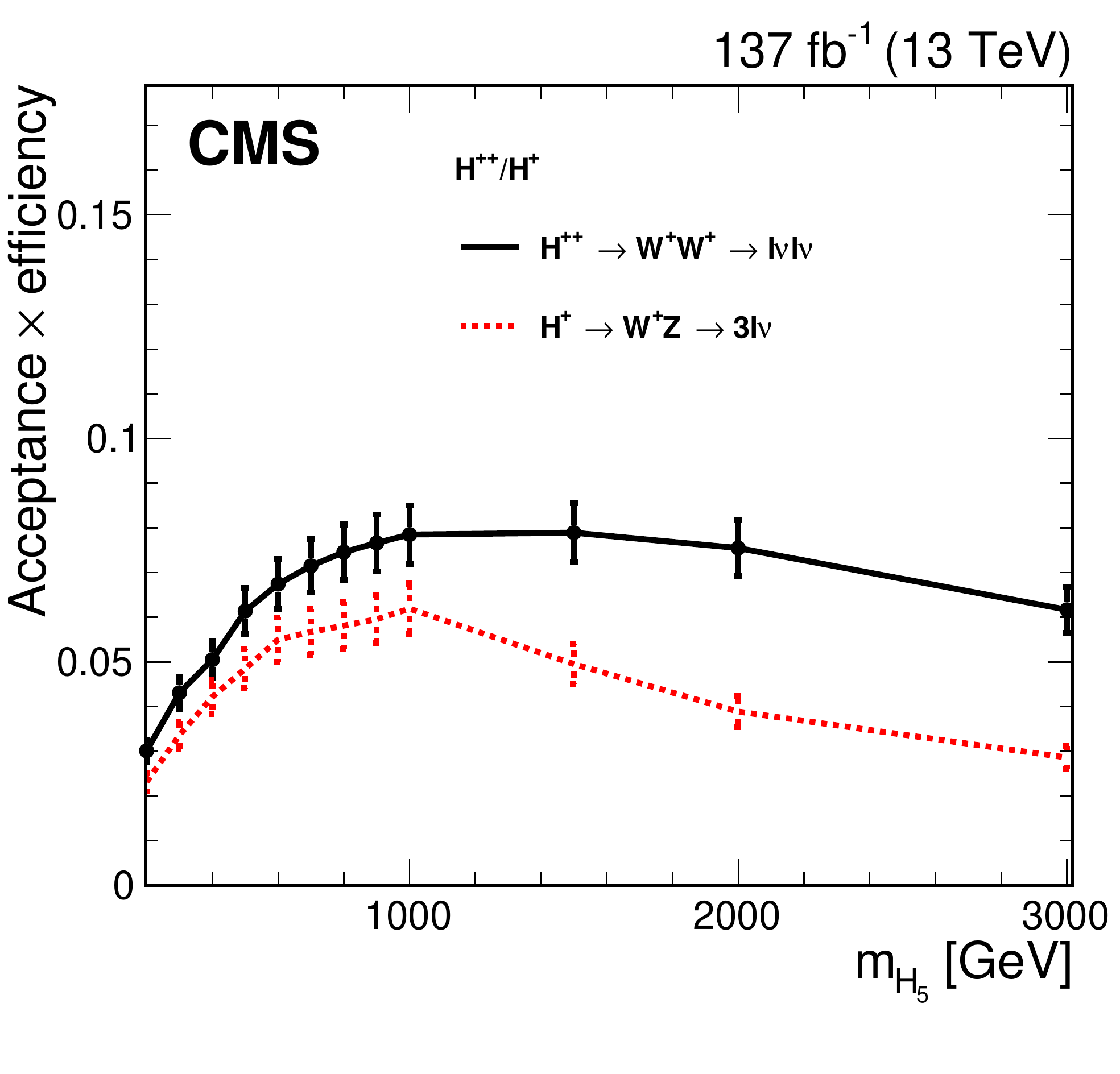}
\caption{The product of acceptance and selection efficiency within the fiducial region 
for the VBF $\PHpmpm \to \WW \to 2\ell 2\nu$ and $\PHpm \to \WZ \to 3\ell\nu$ processes, as a function of $m_{\PH_{5}}$. 
The combination of the statistical and systematic uncertainties is shown. 
The theoretical uncertainties in the acceptance are also included. 
\label{fig:ana_hpp_hp_eff}}
\end{center}
\end{figure}

The distributions of $\mjj$ and $\mT^{\PV\PV}$ in the $\PW\PW$ and $\PW\PZ$ SRs are shown in Fig.~\ref{fig:ssww_mtvv}. The $\mjj$ distributions in the $\PW\PW$ and $\PW\PZ$ SRs are shown with finer binning compared to the binning used in the two-dimensional distribution in the fit.  
Distributions for signal, backgrounds, and data for the bins used in the simultaneous fit are shown in Fig.~\ref{fig:ssww_datacard}. 
The data yields, together with the background expectations with the best fit normalizations for the background-only hypothesis, \ie, 
assuming no contributions from the  $\PHpm$ and $\PHpmpm$ processes, are shown in Table~\ref{tab:yields}. The product of 
kinematic acceptance and selection efficiency within the fiducial region  for the $\PHpmpm \to \WW \to 2\ell 2\nu$ and 
$\PHpm \to \WZ \to 3\ell\nu$ processes, as a function of $m_{\PH_{5}}$, is shown in Fig.~\ref{fig:ana_hpp_hp_eff}. The drop of selection efficiency for the $\PHpm \to \WZ \to 3\ell\nu$ process for masses above $1000\GeV$ is coming from the lepton isolation requirement as the leptons from high-momentum $\PZ$ boson decay are produced with a small angular separation.

\begin{table*}[htbp]
\caption{Expected signal and background yields from various SM processes and observed data events in all regions used in the analysis. The expected background yields are shown with their normalizations from the simultaneous fit for the background-only hypothesis, \ie, assuming no contributions from the  $\PHpm$ and $\PHpmpm$ processes. The expected signal yields are shown for $s_{\PH}=1.0$ in the GM model. The combination of the statistical and systematic uncertainties is shown. \label{tab:yields}}
\begin{center}
{
\newcolumntype{x}{D{,}{\,\pm\,}{3.3}}
\begin{tabular}{lxxxxx{c}@{\hspace*{5pt}}x}
\hline
Process                             &     \multicolumn{1}{c}{$\PW\PW$ SR} & \multicolumn{1}{c}{$\PW\PZ$ SR} &   \multicolumn{1}{c}{Nonprompt CR} & \multicolumn{1}{c}{$\tZq$ CR} & \multicolumn{1}{c}{$\PZ\PZ$ CR} \\
\hline
$\PHpmpm(500) \to \WW$              &   666   ,  68   &	    \NA        &    48.9 ,   5.1 &  	\NA	    &      \NA	      \\
$\PHpm(500) \to \WZ$                &    19.2 ,   2.4 &  107   ,  11   &     1.7 ,   0.2 &     8.0 ,   0.9  &      \NA	      \\[\cmsTabSkip]

$\WW$                               &   230   ,  16   &	    \NA        &    28.2 ,   1.8 &      \NA	    &      \NA	      \\  
$\PW\PZ$               	    &    67.8 ,   5.8 &  196   ,  15   &    10.3 ,   1.0 &    27.2 ,   2.4  &      \NA	      \\
$\PZ\PZ$                  	    &     0.7 ,   0.2 &    6.4 ,   2.0 &     0.1 ,   0.1 &     1.1 ,   0.3  &	 13.3 ,   4.0 \\
Nonprompt                 	    &   262   ,  36   &   22.3 ,   7.7 &   263   ,  21   &     8.4 ,   3.1  &	  0.2 ,   0.2 \\
 $\tVx$               	            &     8.4 ,   1.9 &   17.7 ,   3.3 &    28.8 ,   5.6 &    62   ,  11    &	  0.2 ,   0.1 \\
Other background               	    &    31.1 ,   7.3 &    6.8 ,   1.4 &    21.1 ,   4.2 &     2.2 ,   0.4  &	  0.3 ,   0.1 \\[\cmsTabSkip]
			   	
Total background                    &   600   ,  40   &  249   ,  18   &   352   ,  22   &   101   ,  12    &	 14.0 ,   4.0 \\[\cmsTabSkip]  

Data                                &   602           &   249	       &   352           &   101            &	 14           \\
\hline
\end{tabular}
}
\end{center}
\end{table*}

No significant excess of events above the expectation from the SM background predictions is found. 
The 95\% $\CL$ upper limits on the charged Higgs production cross sections are 
calculated using the modified frequentist approach with the $\CLs$ criterion~\cite{Junk:1999kv,Read1} 
and asymptotic method for the test statistic~\cite{LHC-HCG-Report,Cowan:2010js}.

\begin{figure*}[hbtp]
\centering
\includegraphics[width=0.49\textwidth]{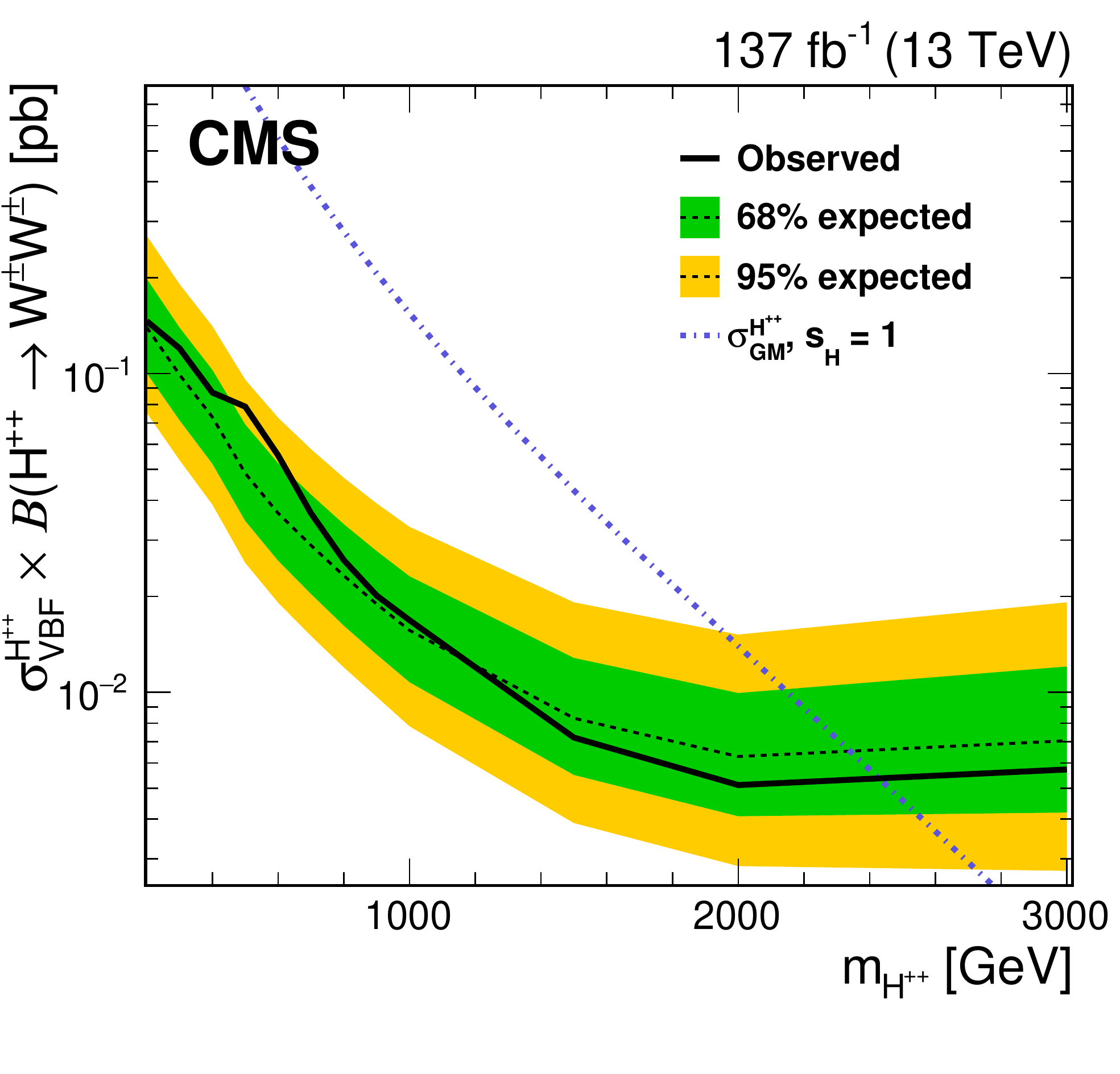}
\includegraphics[width=0.49\textwidth]{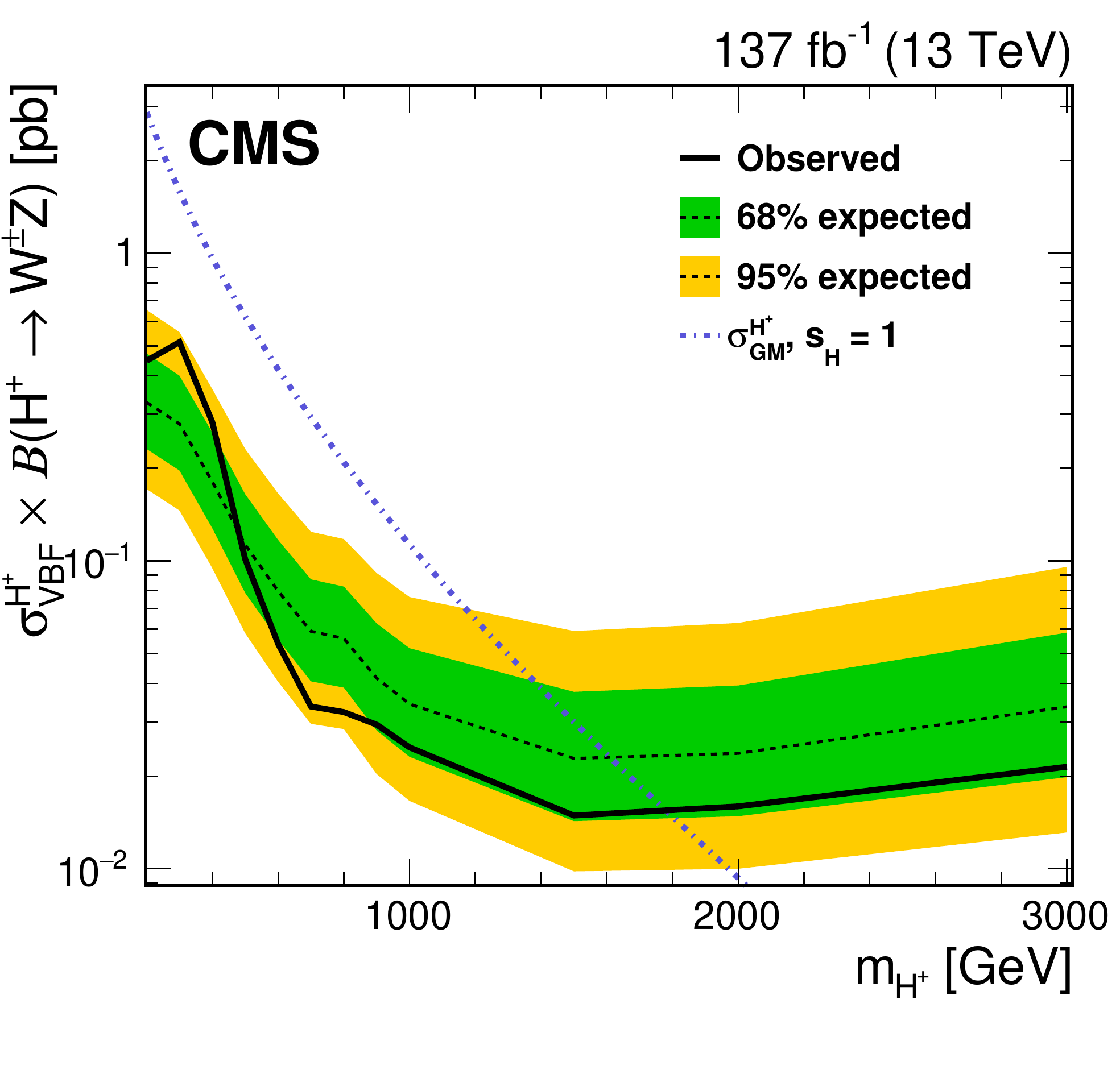}
\includegraphics[width=0.49\textwidth]{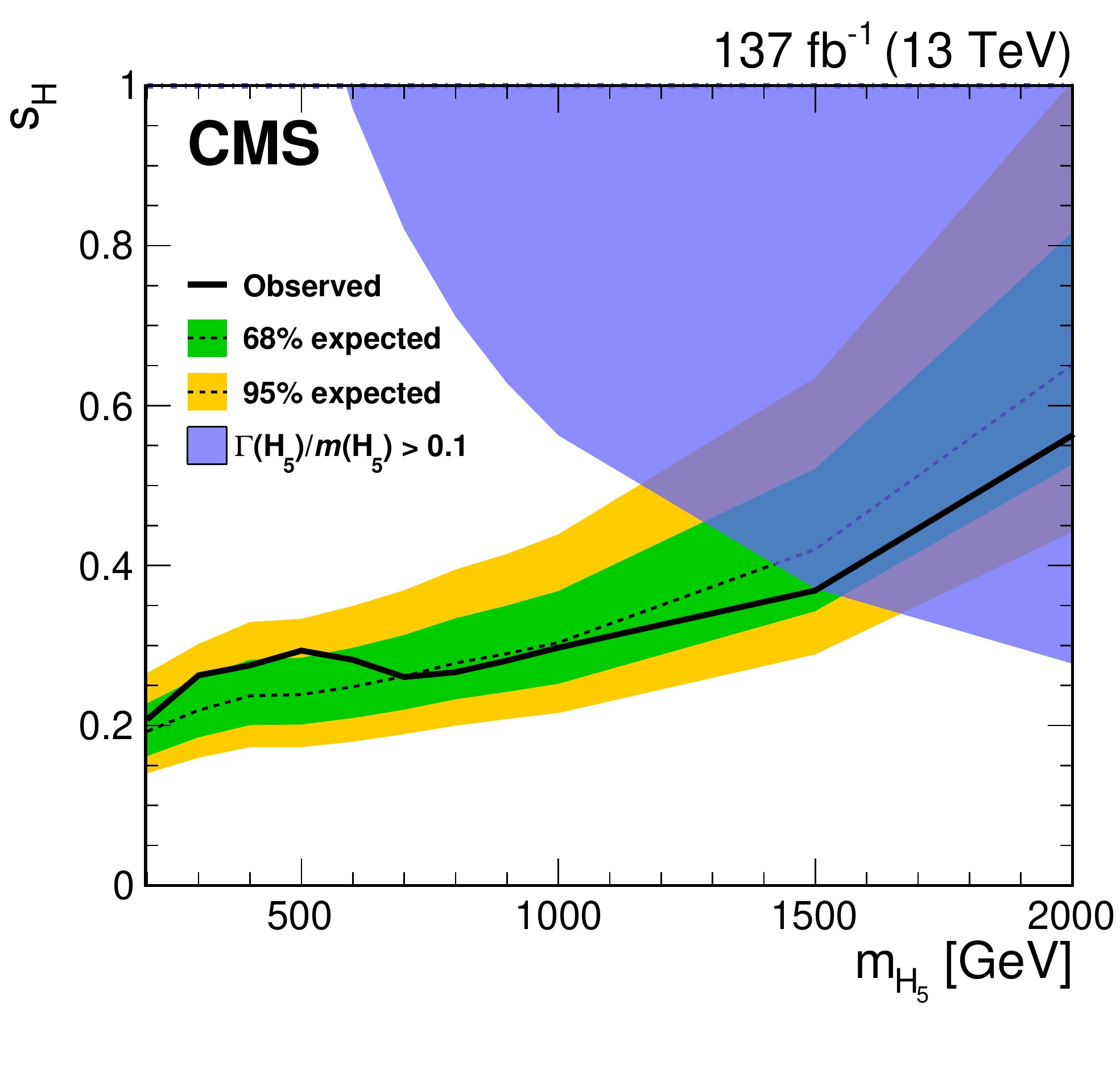}
 \caption{
 Expected and observed exclusion limits at 95\% \CL 
 for $\sigma_\mathrm{VBF}(\PHpmpm) \, \mathcal{B}(\PHpmpm\to \WW)$ as functions of $m_{\PHpmpm}$ (upper left), 
 for $\sigma_\mathrm{VBF}(\PHpm) \, \mathcal{B}(\PHpm\to \WZ)$ as functions of $m_{\PHpm}$ (upper right), 
 and for $s_{\PH}$ as functions of $m_{\PH_{5}}$ in the GM model (lower). The contribution of the $\PHpm$ ($\PHpmpm$) 
 boson signal is set to zero for the derivation of the exclusion limits on the 
 $\sigma_\mathrm{VBF}(\PHpmpm) \, \mathcal{B}(\PHpmpm\to \WW)$  ($\sigma_\mathrm{VBF}(\PHpm) \, \mathcal{B}(\PHpm\to \WZ)$). 
 The exclusion limits for $s_{\PH}$ are shown up to $m_{\PH_{5}}=2000\GeV$, given the low sensitivity 
 in the GM model for values above that mass. Values above the curves are excluded.\label{fig:limits}}
\end{figure*}

{\tolerance=800 Constraints on resonant charged Higgs boson production are derived. The exclusion 
limits on the product of the doubly charged Higgs boson cross section and branching fraction 
$\sigma_\mathrm{VBF}(\PHpmpm) \, \mathcal{B}(\PHpmpm\to \WW)$ at 95\% \CL 
as a function of $m_{\PHpmpm}$ are shown in Fig.~\ref{fig:limits} (upper left). 
The exclusion limits on the product of the charged Higgs boson cross section and branching fraction 
$\sigma_\mathrm{VBF}(\PHpm) \, \mathcal{B}(\PHpm\to \WZ)$ at 95\% \CL 
as a function of $m_{\PHpm}$ are shown in Fig.~\ref{fig:limits} (upper right). The contributions of the $\PHpm$ and $\PHpmpm$ boson signals are set to zero for the derivation of the individual exclusion limits on 
$\sigma_\mathrm{VBF}(\PHpmpm) \, \mathcal{B}(\PHpmpm\to \WW)$  and $\sigma_\mathrm{VBF}(\PHpm) \, \mathcal{B}(\PHpm\to \WZ)$, respectively. The results assume that the intrinsic width of the $\PHpm$ ($\PHpmpm$) boson is 
$\lesssim0.05m_{\PHpm}$ (0.05$m_{\PHpmpm}$), which is below the experimental resolution in the 
phase space considered. The results are also interpreted in the context of the GM model including the simultaneous contributions of the $\PHpm$ and $\PHpmpm$ bosons. The predicted cross sections of the $\PHpm$ and $\PHpmpm$ bosons  at NNLO accuracy in the GM model~\cite{Zaro:2002500} are used for given GM parameter values of $s_{\PH}$ and $m_{\PH_{5}}$. The excluded $s_{\PH}$ values as a function of $m_{\PH_{5}}$ are shown in Fig.~\ref{fig:limits} (lower). 
The blue shaded region shows the parameter space for which the $\PH_{5}$ total width 
exceeds 10\% of $m(\PH_{5})$, where the model is not applicable because of 
perturbativity and vacuum stability requirements~\cite{Zaro:2002500}. 
For the probed parameter space and $\mT^{\PV\PV}$ distribution used for signal 
extraction, the varying width as a function of $s_{\PH}$ is assumed to have 
negligible effect on the result. The observed limit excludes $s_{\PH}$ values 
greater than 0.20--0.35 for the $m_{\PH_{5}}$ range from $200$ to $1500\GeV$. 
The limit improves the sensitivity of the previous CMS results at $13\TeV$, 
where $s_{\PH}$ values greater than about 0.4 and 0.5 are excluded using the 
leptonic decay mode of the 
$\sigma_\mathrm{VBF}(\PHpmpm) \, \mathcal{B}(\PHpmpm\to \WW)$~\cite{Sirunyan:2017ret} and 
$\sigma_\mathrm{VBF}(\PHpm) \, \mathcal{B}(\PHpm\to \WZ)$~\cite{Sirunyan:2019ksz} 
processes, respectively, for the $m_{\PH_{5}}$ range from $200$ to $1000\GeV$. 
Tabulated results are available in the HepData database~\cite{hepdata}.\par}

\ifthenelse{\boolean{cms@external}}{\clearpage}{}

\section{Summary}
\label{sec:summary}
A search for charged Higgs bosons produced in vector boson fusion processes and decaying into vector bosons, using proton-proton collisions at $\sqrt{s}=13\TeV$ at the LHC, is reported. The data sample corresponds 
to an integrated luminosity of 137\fbinv, collected 
with the CMS detector between 2016 and 2018. The search is performed in the leptonic 
decay modes $\WW \to \ell^\pm\PGn\ell'^\pm\PGn$ and $\PW^\pm\PZ \to \ell^\pm\PGn\ell'^\pm\ell'^\mp$, where $\ell, \ell' = \Pe$, $\PGm$. The $\WW$ and $\WZ$ channels 
are simultaneously studied by performing a binned maximum-likelihood fit using the transverse mass $\mT$ and dijet invariant mass $\mjj$ distributions. 
No excess of events with respect to the standard model background predictions is observed. Model independent upper limits at 95\% confidence level are reported on the product of the cross section and branching fraction for vector boson fusion production 
of charged Higgs bosons decaying into vector bosons as a function of mass from 
200 to $3000\GeV$. The results are interpreted in the Georgi--Machacek (GM) model for which the most stringent limits to date are derived. The observed 95\% confidence level limits exclude GM $s_{\PH}$ parameter values greater than 0.20--0.35 for the mass range from $200$ to $1500\GeV$.

\begin{acknowledgments}
  We congratulate our colleagues in the CERN accelerator departments for the excellent performance of the LHC and thank the technical and administrative staffs at CERN and at other CMS institutes for their contributions to the success of the CMS effort. In addition, we gratefully acknowledge the computing centers and personnel of the Worldwide LHC Computing Grid and other centres for delivering so effectively the computing infrastructure essential to our analyses. Finally, we acknowledge the enduring support for the construction and operation of the LHC, the CMS detector, and the supporting computing infrastructure provided by the following funding agencies: BMBWF and FWF (Austria); FNRS and FWO (Belgium); CNPq, CAPES, FAPERJ, FAPERGS, and FAPESP (Brazil); MES (Bulgaria); CERN; CAS, MoST, and NSFC (China); MINCIENCIAS (Colombia); MSES and CSF (Croatia); RIF (Cyprus); SENESCYT (Ecuador); MoER, ERC PUT and ERDF (Estonia); Academy of Finland, MEC, and HIP (Finland); CEA and CNRS/IN2P3 (France); BMBF, DFG, and HGF (Germany); GSRT (Greece); NKFIA (Hungary); DAE and DST (India); IPM (Iran); SFI (Ireland); INFN (Italy); MSIP and NRF (Republic of Korea); MES (Latvia); LAS (Lithuania); MOE and UM (Malaysia); BUAP, CINVESTAV, CONACYT, LNS, SEP, and UASLP-FAI (Mexico); MOS (Montenegro); MBIE (New Zealand); PAEC (Pakistan); MSHE and NSC (Poland); FCT (Portugal); JINR (Dubna); MON, RosAtom, RAS, RFBR, and NRC KI (Russia); MESTD (Serbia); SEIDI, CPAN, PCTI, and FEDER (Spain); MOSTR (Sri Lanka); Swiss Funding Agencies (Switzerland); MST (Taipei); ThEPCenter, IPST, STAR, and NSTDA (Thailand); TUBITAK and TAEK (Turkey); NASU (Ukraine); STFC (United Kingdom); DOE and NSF (USA).
    
  \hyphenation{Rachada-pisek} Individuals have received support from the Marie-Curie program and the European Research Council and Horizon 2020 Grant, contract Nos.\ 675440, 724704, 752730, 765710 and 824093 (European Union); the Leventis Foundation; the Alfred P.\ Sloan Foundation; the Alexander von Humboldt Foundation; the Belgian Federal Science Policy Office; the Fonds pour la Formation \`a la Recherche dans l'Industrie et dans l'Agriculture (FRIA-Belgium); the Agentschap voor Innovatie door Wetenschap en Technologie (IWT-Belgium); the F.R.S.-FNRS and FWO (Belgium) under the ``Excellence of Science -- EOS" -- be.h project n.\ 30820817; the Beijing Municipal Science \& Technology Commission, No. Z191100007219010; the Ministry of Education, Youth and Sports (MEYS) of the Czech Republic; the Deutsche Forschungsgemeinschaft (DFG), under Germany's Excellence Strategy -- EXC 2121 ``Quantum Universe" -- 390833306, and under project number 400140256 - GRK2497; the Lend\"ulet (``Momentum") Program and the J\'anos Bolyai Research Scholarship of the Hungarian Academy of Sciences, the New National Excellence Program \'UNKP, the NKFIA research grants 123842, 123959, 124845, 124850, 125105, 128713, 128786, and 129058 (Hungary); the Council of Science and Industrial Research, India; the Ministry of Science and Higher Education and the National Science Center, contracts Opus 2014/15/B/ST2/03998 and 2015/19/B/ST2/02861 (Poland); the National Priorities Research Program by Qatar National Research Fund; the Ministry of Science and Higher Education, project no. 0723-2020-0041 (Russia); the Programa Estatal de Fomento de la Investigaci{\'o}n Cient{\'i}fica y T{\'e}cnica de Excelencia Mar\'{\i}a de Maeztu, grant MDM-2015-0509 and the Programa Severo Ochoa del Principado de Asturias; the Thalis and Aristeia programs cofinanced by EU-ESF and the Greek NSRF; the Rachadapisek Sompot Fund for Postdoctoral Fellowship, Chulalongkorn University and the Chulalongkorn Academic into Its 2nd Century Project Advancement Project (Thailand); the Kavli Foundation; the Nvidia Corporation; the SuperMicro Corporation; the Welch Foundation, contract C-1845; and the Weston Havens Foundation (USA).
\end{acknowledgments}

\bibliography{auto_generated}
\cleardoublepage \appendix\section{The CMS Collaboration \label{app:collab}}\begin{sloppypar}\hyphenpenalty=5000\widowpenalty=500\clubpenalty=5000\vskip\cmsinstskip
\textbf{Yerevan Physics Institute, Yerevan, Armenia}\\*[0pt]
A.M.~Sirunyan$^{\textrm{\dag}}$, A.~Tumasyan
\vskip\cmsinstskip
\textbf{Institut f\"{u}r Hochenergiephysik, Wien, Austria}\\*[0pt]
W.~Adam, J.W.~Andrejkovic, T.~Bergauer, S.~Chatterjee, M.~Dragicevic, A.~Escalante~Del~Valle, R.~Fr\"{u}hwirth\cmsAuthorMark{1}, M.~Jeitler\cmsAuthorMark{1}, N.~Krammer, L.~Lechner, D.~Liko, I.~Mikulec, P.~Paulitsch, F.M.~Pitters, J.~Schieck\cmsAuthorMark{1}, R.~Sch\"{o}fbeck, M.~Spanring, S.~Templ, W.~Waltenberger, C.-E.~Wulz\cmsAuthorMark{1}
% ...
\vskip\cmsinstskip
\dag: Deceased\\
1:  Also at Vienna University of Technology, Vienna, Austria\\
% ...
\end{sloppypar}
\end{document}